\documentclass[11pt,a4paper]{article}
\usepackage{graphicx} 
\usepackage{subcaption}
\usepackage{hyperref}
\usepackage{amsmath}
\usepackage[a4paper, left=1in, right=1in]{geometry}
\usepackage{xcolor}
\usepackage[capitalise]{cleveref}

\usepackage{authblk}
\usepackage[numbers,sort&compress]{natbib}
\hypersetup{
    colorlinks=true,
    linkcolor=blue,
    filecolor=magenta,      
    urlcolor=blue,
    citecolor=blue,
    pdfpagemode=FullScreen,
    }

\title{\textbf{BSM Searches at a Photon Collider with Energy $E_{\gamma\gamma}< 12$~GeV}}

\author[1]{Marten Berger\footnote{marten.berger@desy.de}}
\author[1,2]{Gudrid Moortgat-Pick\footnote{gudrid.moortgat-pick@desy.de}}

\affil[1]{II. Institut für Theoretische Physik, Universität Hamburg, Luruper Chaussee 149, 22761 Hamburg, Germany}
\affil[2]{Deutsches Elektronen-Synchrotron DESY, Notkestr.~85, 22607 Hamburg, Germany}
\date{}

\begin{document}

\makeatletter
\def\ps@firstpage{%
  \def\@oddhead{\hfill\small DESY-26-029}%
  \def\@evenhead{}%
  \def\@oddfoot{}%
  \def\@evenfoot{}%
}
\makeatother

\maketitle
\thispagestyle{firstpage}

\begin{abstract}
The possibility of a photon collider extension to the beam dump of the $17.5$ GeV European XFEL has already been discussed before as the first high energy collider of its sort. It would not only be the first proof of concept and test of a photon collider but would also be a collider without competition in the region of $E_{\gamma\gamma}=5-12$ GeV for photon-photon collision. In this range, $b\bar{b}$ and $c\bar{c}$ resonances, tetraquarks and mesonic molecules can be observed. Furthermore, some BSM processes can also be reached in this range. In this paper we want to discuss the possibility of observing ALPs in the process of light-by-light scattering at such a collider. We will use a simplified description of the Compton backscattering process to get a first look at cross sections for the Standard Model light-by-light scattering and the extension including ALPs. Furthermore, we extend this to the full beam dynamics included prediction, discuss all effects that are important when working with a photon collider and show that the photon collider with energy $E_{\gamma\gamma}<12$ GeV would offer an extended physics reach compared to current limits.
\end{abstract}

\section{Introduction}

Future linear colliders~\cite{ECFADESYLCPhysicsWorkingGroup:2001igx,LinearColliderACFAWorkingGroup:2001awr,LinearColliderAmericanWorkingGroup:2001tzv,Abe:2001rdr,Abe:2001grn,LinearColliderAmericanWorkingGroup:2001rzk} are the promising next high-energy machines in order to tackle the most pressing open questions in particle physics. They offer several unique advantages, e.g. tunable energies and polarised beams~\cite{Moortgat-Pick:2005jsx} and complement current collider projects in exploring the Standard Model of particle physics (SM) and beyond (BSM) e.g. \cite{LinearColliderVision:2025hlt,Moortgat-Pick:2015lbx,LHCLCStudyGroup:2004iyd} and references therein).
One of the primary urgent topics for current and future research is the study of the Higgs boson properties, following its discovery at the LHC in 2012 \cite{ATLAS_2012,CMS_2012}. Linear colliders, such as the International Linear Collider (ILC) \cite{Bambade:2019fyw}, the Hybrid, Asymmetric, Linear Higgs Factory (HALHF) \cite{HALHF_2023} or the Compact Linear Collider (CLIC) \cite{CLIC_2016}, are designed to operate mainly at center-of-mass energies between 250 GeV and 3 TeV. This energy range is, for instance, ideal for producing Higgs bosons and studying their properties, including the trilinear Higgs coupling (cms $\ge 550$ GeV required for Higgs pair production), which is essential for probing the structure of the Higgs potential \cite{deBlas:2019rxi, Bechtle:2024acc, LinearColliderVision:2025hlt}. The extremely precisely known initial state and low background environment at $e^+e^-$-colliders would allow high precision measurements of the Higgs properties and reveal subtle deviations from the SM, opening new windows for detecting BSM physics.\\
At linear colliders, the beam can also be used to obtain high energy photons via Compton backscattering of laser light from the electron beam, giving access to $\gamma\gamma$ and $\gamma e$ interactions with luminosities and energies comparable to that of the $e^+e^-$ mode \cite{Ginzburg1,GINZBURG2,Ginzburg3,Telnov_1989}. Such options would extend the physics reach and perfectly complement the searches in the $e^+e^-$-mode, offering unique access to so far inaccessible BSM or very rarely observed phenomena of the SM \cite{Berger:2025ijd,ourpaper, Angel:2026hfp}. The photon collider is therefore an ideal extension in particular of all and future $e^+e^-$-linear-colliders. Since all of them are still in the planning phase, it is advantageous to include a gamma-gamma option already in the current design phase. Furthermore, it would even be advantageous to apply this technology already at current devices in order to be maximize their physical potential and test this technology. From accelerator side a lot of work has been done, for instance establishing powerful Monte-Carlo simulation tools such as \texttt{CAIN} \cite{CHEN1995107}, that can calculate a realistic energy and photon luminosity spectrum for a given collider. \\
The European-XFEL is a free-electron laser that has been successfully in operation since 2017, generating high energy X-ray beams by using a $17.5$ GeV electron linac. The electron beam is sent to the beam dump after passing through the undulators. For the photon collider, however, the used electron beam is split and redirected along two curves and collides, thus create a $35$ GeV $e^-e^-$-collider and allowing the creation of high-energy photons via Compton backscattering, i.e. a photon collider with collisions at $E_{\gamma\gamma}=12$ GeV. This idea has been proposed by~\cite{TELNOV_talk, Telnov_2020},  allowing physics projects and searches for heavy two-photon resonances even at low GeV, e.g.~\cite{Beloborodov:2022byx}.
The proposal uses the same laser system that could also be implemented at $100-1000$ GeV colliders for backscattering.
At future high-energy linear colliders a photon collider option could reach up to $80 \%$ of the $e$-beam energy \cite{TESLA_TDR, Bambade:2019fyw} when using optical lasers, or even $\sim100 \%$ when using an XFEL-like laser \cite{Barklow:2024XCC} for the Compton backscattering. Even studying single Higgs and Higgs-pair production at photon colliders with cms $\sim125$ GeV and $280$ GeV respectively would be possible, allowing unique searches for exploring of Higgs physics \cite{Berger:2025ijd, ourpaper, Castelazo:2026iuu}.\\ 
A first natural process to study at photon colliders is light-by-light (LbyL) scattering. It was first proposed in the 1930's by Euler \cite{Euler:1935qgl}. LbyL-scattering has been worked on over many years, with great progress being made to calculate this process within QED \cite{Karplus:1950zz, DeTollis:1964una, DeTollis:1965vna}. Later the calculations were extended to the electroweak SM, including the $W$-loop contributions \cite{Jiang:1992sm, Dong:1992fa, Dong:1992hg}. The first direct observation of light-by-light scattering has been achieved by ATLAS in 2019, by colliding heavy lead ions \cite{ATLAS:2019azn}, following the proposal of \cite{dEnterria:2013zqi}. A true photon collider would give unprecedented direct access to this process, allowing to exploit LbyL-scattering for BSM searches.\\
Axion-like particles (ALPs) could cause deviations to SM expectations. They are hypothetical, weakly interacting BSM particles, initially inspired by the axion , which was proposed to solve the strong CP problem in quantum chromodynamics (QCD) \cite{PQ_1977}. ALPs generalize the idea by decoupling the mass and coupling of these particles from the QCD context. ALPs are typically described as pseudo-Nambu-Goldstone bosons arising from spontaneous symmetry breaking at high energy scales. Their weak interactions with photons and other SM particles turns them into ideal candidates for dark matter, especially in the context of light, sub-eV masses, where they could form coherent, cold dark matter states. Many ongoing experiments, such as CAST \cite{Iguaz:2011iz}, ADMX \cite{Khatiwada_2021}, and IAXO \cite{IAXO:2025ltd}, aim to detect very light axions with masses $m_a<1 \text{ eV}$, but there are also collider searches and proposals ongoing for heavier ALPs in the keV-TeV range \cite{AxionPortal, ColliderALPs, Ma:2025ymt, Baldenegro:2018hng}, for example at ATLAS \cite{Knapen:2017ebd}, CMS \cite{CMS:2018erd,CMS:2024bnt} and Belle II \cite{Dolan_2017}. So far the collider-based searches are using only virtual photons that are radiated off the electron / heavy-ions. A photon collider with real photons, however, would be unique in its access to both the light-by-light scattering as well as the ALP induced process $\gamma\gamma\rightarrow a\rightarrow\gamma\gamma$, allowing complementary and precision searches for the SM and its deviations, i.e. hints of BSM physics. There have also been a few recent studies looking at the potential of a photon collider for energies above 1 TeV \cite{Inan:2020aal} using the analytical equations given by \cite{Ginzburg3}, as well as laser-assisted light-by-light, colliding a high-energy gamma-ray beam with a laser pulse \cite{Ma:2025ymt}. However, no studies have applied a real photon spectrum from Monte-Carlo simulation programs such as \texttt{CAIN} so far, incorporating many of the collider parameters.\\
In this work we will discuss the photon collider and compare both methods for obtaining the photon spectrum, the analytical method and the full \texttt{CAIN} simulation. Furthermore, we will study the SM light-by-light scattering process, including an additional ALP, and focus on parameters accessible for the photon collider based on the European-XFEL.\\
In Sec. \ref{sec:GammaGamma} we will discuss the concepts and provide the important equations for a photon collider and discuss the energy spectra and luminosities for different beam set-ups and simulation methods. In Sec. \ref{sec:lbyl} the SM light-by-light scattering process at the photon collider is discussed and is extended in Sec. \ref{sec:alp} to the inclusion of an additional ALP. The conclusions on such a low-energy photon collider at the European-XFEL are given in Sec. \ref{sec:conclusion}.

\section{Introduction to Photon collider}
\label{sec:GammaGamma}

\begin{figure}[htbp]
    \centering
    \includegraphics[scale=0.3]{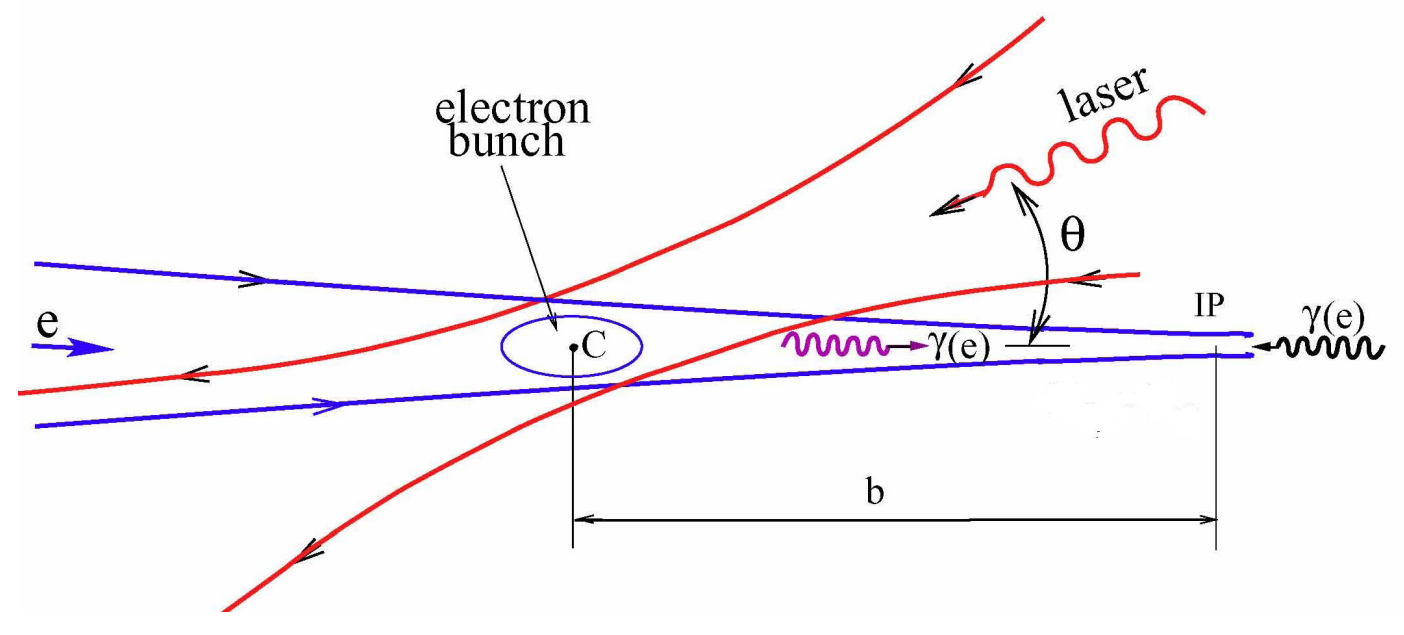}
    \caption{General scheme of a $\gamma\gamma$, $\gamma e$ collider \cite{Telnov_2020}.}
    \label{fig:scheme}
\end{figure}

A photon collider uses Compton backscattering in order to convert high-energy electrons into high-energy photons. Laser photons with energy $\omega_0$ collide with the electrons at a conversion point ($C$), a short distance $b$ before the photon-photon interaction point ($IP$), as illustrated in Fig. \ref{fig:scheme}. This process results in colliding real photons and expands the physics potential of an $e^+e^-$-collider by $\gamma\gamma$ and $\gamma e$ collisions with luminosities comparable to those of $e^+e^-$ collisions.

The maximum energy of the backscattered high-energy photons $E_{\gamma,\text{max}}$ ($\hbar=1, c=1$) is given by
\begin{equation}
    E_{\gamma,\text{max}} = \frac{x}{x+1}E_0,
    \label{eq:omega}
\end{equation}
where
\begin{equation}
    x = \frac{4E_0\omega_0}{m_e^2}\cos^2\frac{\theta}{2}\simeq  15.3 \left[\frac{E_0}{\text{TeV}}\right]\left[\frac{\omega_0}{\text{eV}}\right]=19\left[\frac{E_0}{\text{TeV}}\right]\left[\frac{\mu\text{m}}{\lambda}\right]
    \label{eq:x}
\end{equation}
with $m_e$ denoting the electron mass, $\theta$ the small collision angle, $E_0$ the electron energy, $\omega_0$ the laser photon frequency and $\lambda$ the laser photon wavelength.
In order to get photons as close as possible to the energy of the electrons, a large value of $x$ is needed. However, there are two important thresholds for $x$ at which many high-energy photons are lost to $e^+e^-$-pair production. The first process, the Breit-Wheeler process $\gamma\gamma_0\rightarrow e^+e^-$ (here $\gamma$ is the scattered photon and $\gamma_0$ is a laser photon), is possible when $E_{\gamma,\text{max}}\omega_0=m_e^2c^4$, corresponding to $x=4.8$; and the second process the Bethe-Heitler process $e^-\gamma\gamma_0\rightarrow e^-e^+e^-$, occurs when $x=8.0$. Therefore, going beyond $x=4.8$ leads to a significant loss in conversion and in luminosity; however, recent papers have proposed the idea of using XFELs instead of optical lasers, resulting in values $x>1000$ while still achieving high conversion rates and luminosities at the peak energy \cite{Barklow:2024XCC}. 

The total cross section for the Compton backscattering, $\sigma_c$, is given by \cite{Ginzburg3}

\begin{equation}
    \sigma_c = \sigma_c^{np}+\lambda_e P_c\sigma_c^1,
    \label{eq:comptonPol}
\end{equation}

\begin{equation}
    \sigma_c^{np} = \frac{2\pi\alpha^2}{x m_e^2}\left[ \left(\ 1-\frac{4}{x}-\frac{8}{x^2}\right) \ln{(x+1)}+\frac{1}{2}+\frac{8}{x}-\frac{1}{2(x+1)^2}\right],
    \label{eq:compton}
\end{equation}
\begin{equation}
    \sigma_c^1=\frac{2\pi\alpha^2}{xm_e^2}\left[ \left( 1+\frac{2}{x} \right)\ln{x+1}-\frac{5}{2}+\frac{1}{x+1}-\frac{1}{2(x+1)^2} \right],
    \label{eq:comptonPolcorrection}
\end{equation}
where $\alpha$ is the fine-structure constant, $\lambda_e$ and $P_c$ are the mean electron and laser photon helicity respectively. The first part $\sigma_c^{np}$ represents the Compton cross section for unpolarised beams, when both initial beams have a nonzero helicity, i.e. for $\lambda_e P_c \neq 0$, the addition of the correction term $\sigma_c^1$ gives the total polarised Compton cross section , cf. Eq. \ref{eq:comptonPol}. 

\begin{figure}[h]
    \centering
    \begin{subfigure}{0.45\linewidth}
        \centering
        \includegraphics[width=1\linewidth]{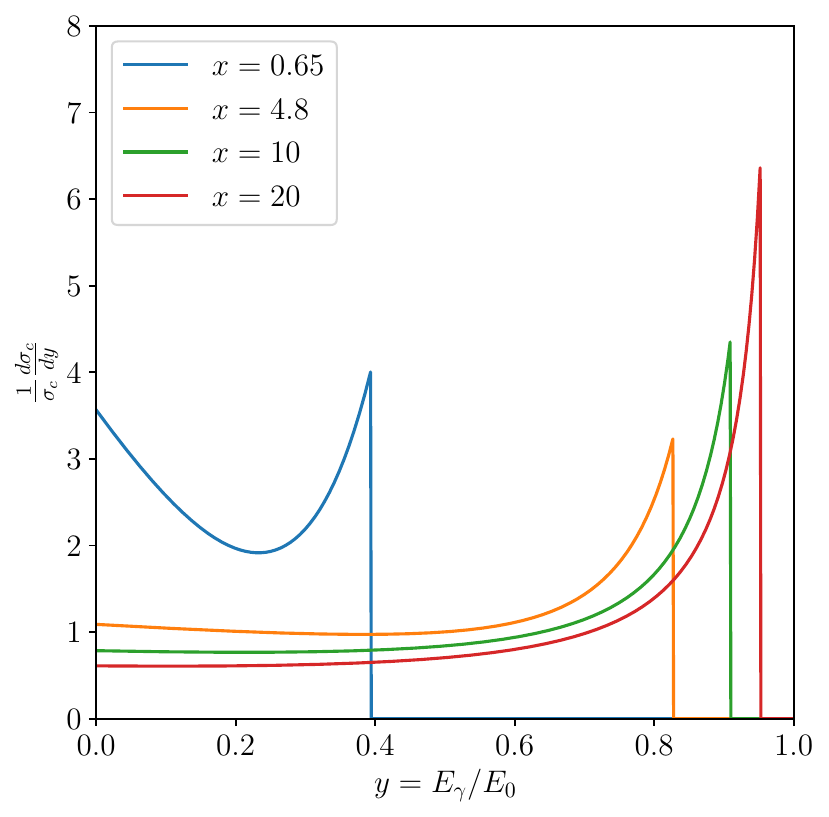}
    \end{subfigure}
    \begin{subfigure}{0.45\linewidth}
        \centering
        \includegraphics[width=1\linewidth]{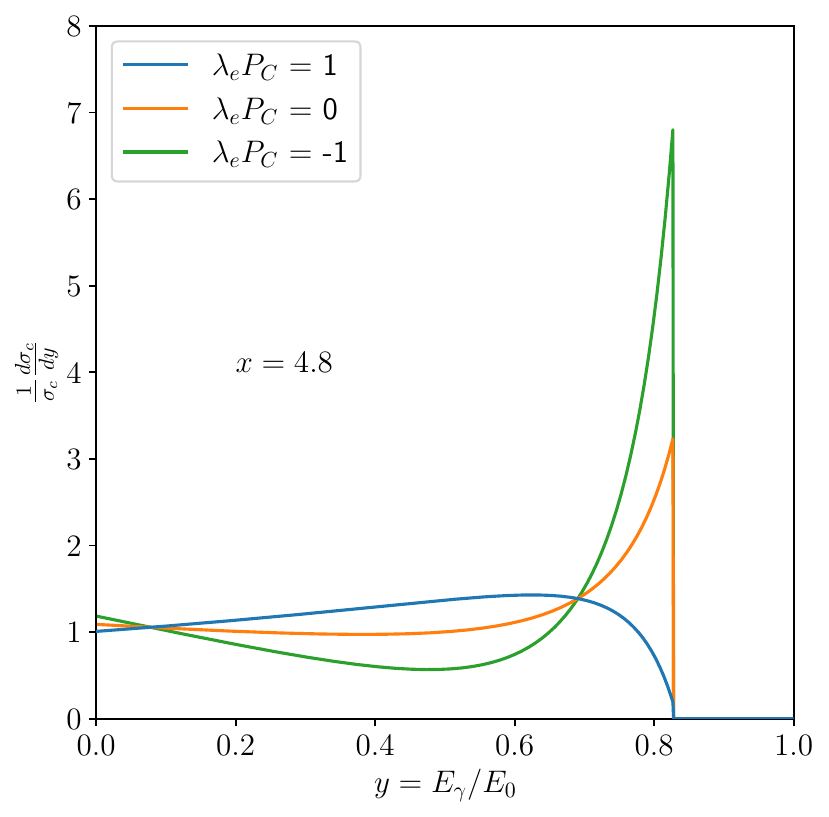}
    \end{subfigure}
    \caption{Energy spectrum for backscattered photons for different values of $x$ (left plot) and different combinations of initial electron and photon polarisation with $x=4.8$ (right plot).}
    \label{fig:fFunc}
\end{figure}

The resulting energy spectrum for the backscattered photons can by found as:
\begin{equation}
    \begin{split}
        \frac{1}{\sigma_c} \frac{d\sigma_c}{dy} &\equiv f(x,y) \\&= \frac{2\pi\alpha^2}{\sigma_c x m_e^2} \left[ \frac{1}{1-y} + 1-y-4r(1-r)-\lambda_e P_c rx(2r-1)(2-y) \right]
    \end{split}
    \label{eq:energySpecPol}
\end{equation}
with
\begin{equation}
    r = \frac{y}{x(1-y)}\leq 1,
\end{equation}
where $y=E_\gamma / E_0$ is the energy fraction of the scattered photon energy $E_\gamma$ normalised to the incoming electron energy $E_0$. The energy spectrum and the effects of different polarisations as well as different values for $x$ can be seen in Fig. \ref{fig:fFunc}. For an initial polarisation of $\lambda_e P_c=-1$ the spectrum has a higher peak at $y=y_{max}$ compared to the unpolarised spectrum. On the other hand for an initial polarisation of $\lambda_e P_c=1$ the spectrum is lower and the maximum is even moved to a point away from $y=y_{max}$. Comparing the energy spectrum for different $x$, it can be seen that the spectrum always follows a similar path but is cut off at different values for $y$.

In the photon-photon mode, when both beams are converted, the luminosity spectrum is given by:
\begin{equation}
    \begin{split}
        \frac{1}{k^2L_{ee}}\frac{\text{d}L_{\gamma\gamma}}{\text{d}z} = &2z\int_{z^2/y_{max}}^{y_{max}}\frac{\text{d}y}{y}f(x,y)f\left(x,\frac{z^2}{y}\right)*\\ *&I_0\left( \rho^2\sqrt{\left( \frac{y_{max}}{y} -1 \right)\left( \frac{y_{max}y}{z^2}-1 \right)} \right)\text{exp}\left[ -\frac{\rho^2}{2}\left( \frac{y_{max}}{y}+\frac{y_{max}y}{z^2}-2 \right) \right]\\
        =&2z\int_{z^2/y_{max}}^{y_{max}}\frac{\text{d}y}{y}f(x,y)f\left(x,\frac{z^2}{y}\right)*F(\rho),
    \end{split}
    \label{eq:lumiGamma}
\end{equation}
\begin{equation} 
    \label{eq:rho}
    \rho^2 = \left(\frac{b}{\gamma\sigma_x}\right)^2+\left(\frac{b}{\gamma\sigma_y}\right)^2,
\end{equation}
where $k$ is the conversion coefficient representing the average number of high energy photons per electron, $L_{ee}$ the geometric luminosity with $L_{\gamma\gamma} = k^2L_{ee}$ and the parameter $\rho$ check $\sigma_{x,y}$ definition, should be beam spread depending on the distance $b$ between the Compton scattering point $C$ and the $IP$ as shown in Fig. \ref{fig:scheme}; in many cases $\rho=0$ is chosen, assuming the conversion occurs directly at the $IP$. To show the most common form with this assumption We have combined the $\rho$ dependency in the function $F(\rho)$ that follows for $\rho\rightarrow0, F(\rho)\rightarrow1$.
\begin{figure}
    \centering
    \begin{subfigure}{0.45\linewidth}
        \centering
        \includegraphics[width=1\linewidth]{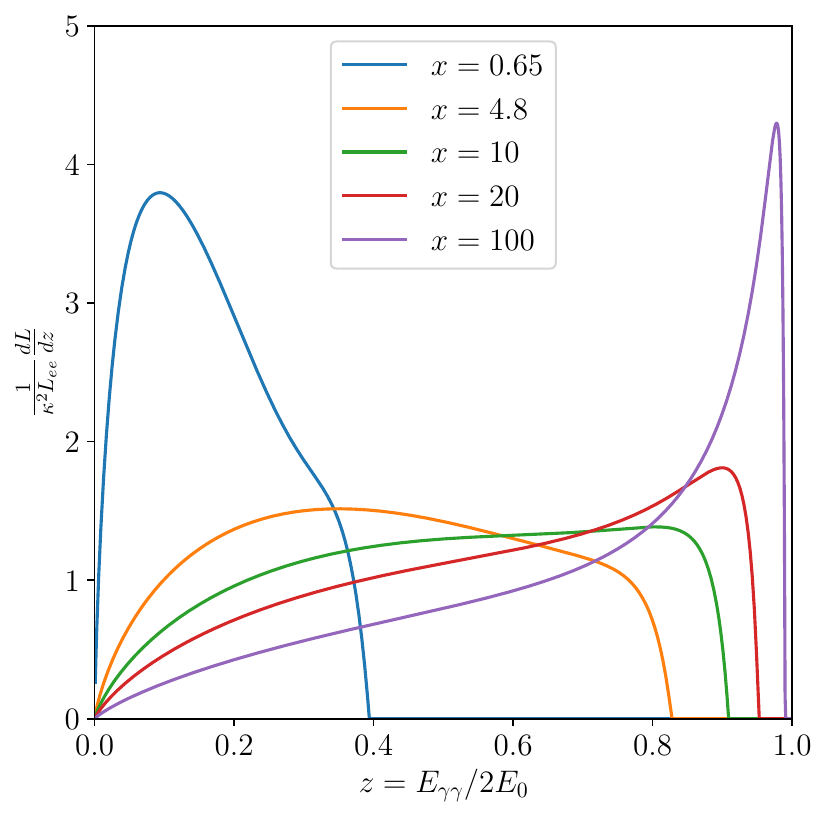}
    \end{subfigure}
    \begin{subfigure}{0.45\linewidth}
        \centering
        \includegraphics[width=1.05\linewidth]{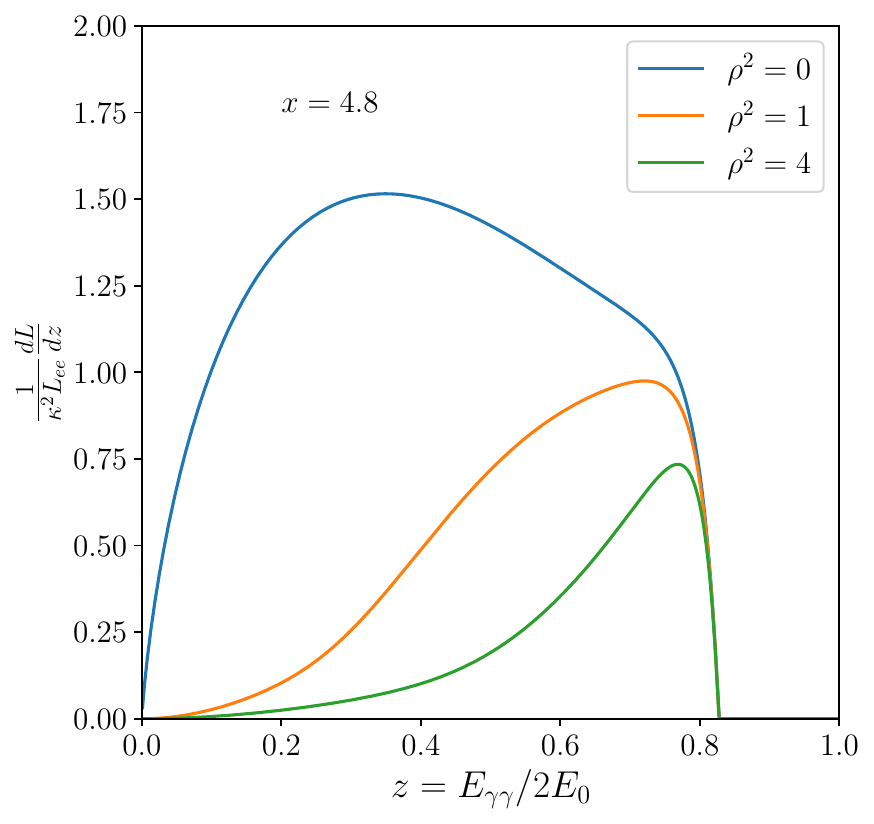}
    \end{subfigure}
    \caption{Luminosity spectra for a photon collider at different values of $x$ (left plot), corresponding to different energies of the $e$-beam and laser (eq. (\ref{eq:x})), and for various distances $b$ between the $C$ and $IP$ (right plot), defined by the parameter $\rho^2$ (eq. (\ref{eq:rho})). Both electron beams and lasers are unpolarised $\lambda_e=P_c=0$.}
    \label{fig:dLdz_unpol}
\end{figure}
The impact of the parameters $x$ and $\rho$ on the luminosity spectrum can be seen in Fig. \ref{fig:dLdz_unpol}. For larger values of $x$ the spectrum becomes broader and reaches higher peak values of $z = E_{\gamma\gamma}/2E_0$, where $E_{\gamma\gamma}$ is the photon-collision energy. At larger values $x>10$ the spectrum starts to peak in the high energy range. Achieving $x=1000$ results in a sharp peak close to its maximum energy, while having very small contributions from lower values of $z$. Getting a more monochromatic beam can also be achieved for optical lasers ($x=4.8$) by increasing $\rho$ as shown in the right plot, but at cost of the overall luminosity: for instance, for $\rho^2=4$ the spectrum is peaked close to $z_{max}$. With different choices of $\rho$, the resulting spectrum can either be more peaked, optimal for total cross section measurements, or broader but an overall larger spectrum is good for resonance studies. These effects are similar for all polarisation configurations. In the left plot of Fig. \ref{fig:dLdz_polarized}, the luminosity spectrum for different choices of polarisations is shown. It can be seen that in order to maximize the spectrum at high values of $z$ the optimal configuration is when both beams have $\lambda_eP_c=\tilde{\lambda}_e\tilde{P}_c=-1$ (green curve). But for $x> 4.8$ the choice of $\lambda_eP_c=\tilde{\lambda}_e\tilde{P}_c=1$ is better since, due to the opposite sign for the helicity products the pair production is suppressed  \cite{Barklow:2024XCC}, and less high-energy photons are lost.
\begin{figure}[htbp]
    \centering
    \begin{subfigure}{0.5\linewidth}
        \centering
        \includegraphics[width=0.9\linewidth]{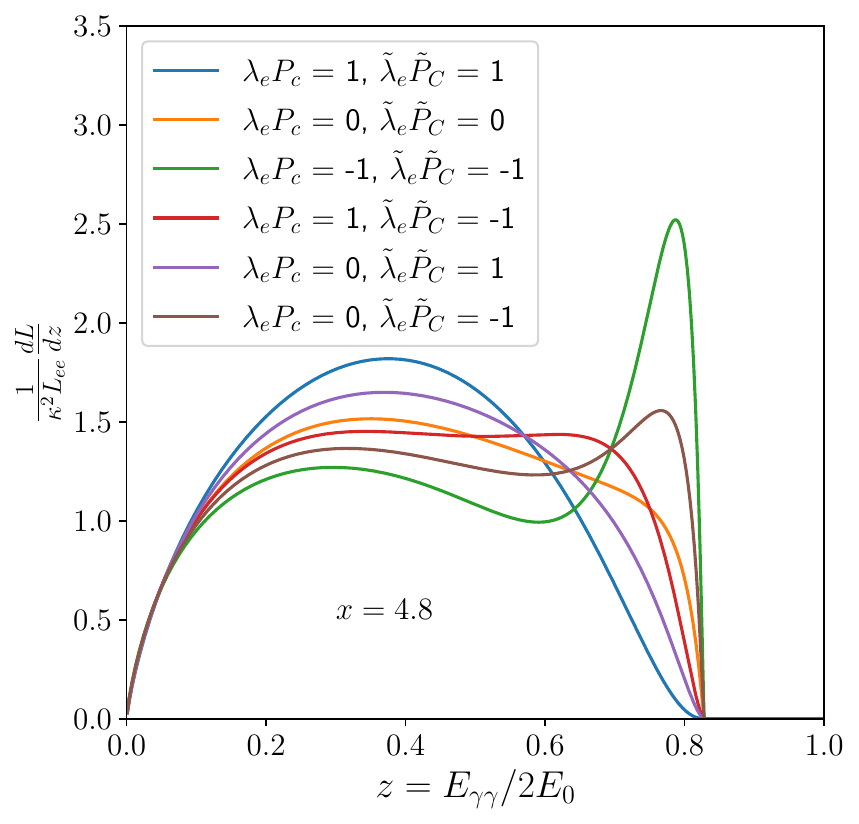}
    \end{subfigure}
    \begin{subfigure}{0.5\linewidth}
        \centering
        \includegraphics[width=0.9\linewidth]{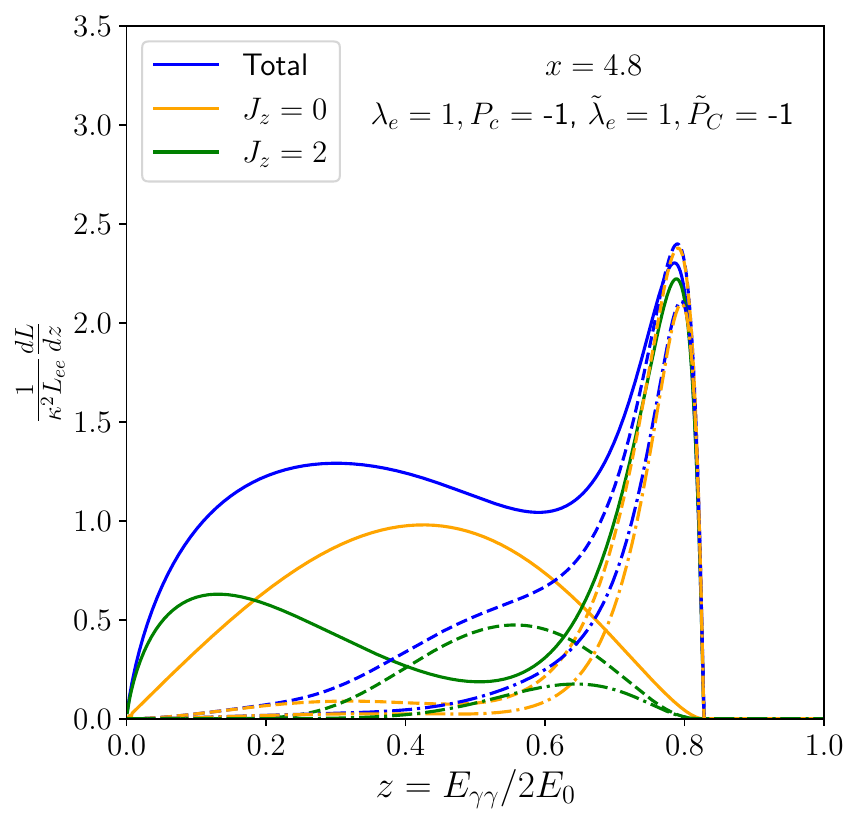}
    \end{subfigure}
    \caption{Luminosity spectrum for different choices of incoming electron and laser photon polarisations (left) and the contribution of the $J_z=0,2$ states for different values of $\rho$ (right). The solid lines are for $\rho^2=0$, the dashed for $\rho^2=1$ and the dotted-dashed for $\rho^2=4$.}
    \label{fig:dLdz_polarized} 
\end{figure}
When including the polarisation of the incoming electrons and laser photons, the high energy photons will also obtain a corresponding polarisation. The mean helicity of the scattered photon is given by \cite{TESLA_TDR}
\begin{equation}
    \langle\lambda_\gamma\rangle = \frac{\lambda_e x r [1 + (1 - y) (2r - 1)^2] - P_c (2r - 1)[(1 - y)^{-1} + 1 - y]}{(1 - y)^{-1} + 1 - y - 4r (1 - r) - \lambda_e P_c x r (2 - y)(2r - 1)}.
    \label{eq:helicityDistribution}
\end{equation}
If the helicity of either the laser photon ($P_c \neq 0$) or the electron ($\lambda_e \neq 0$) is non-zero, the resulting photons will have a non-zero averaged helcity ($\langle\lambda_\gamma\rangle \neq 0$), too. This gives access to polarised high energy photons even when one of the beams is unpolarised ($\lambda_e P_c=0$). The mean helicity of the Compton scattered photon for different initial laser photon and electron helicities is shown in Fig. \ref{fig:helicitiesExample}, with the plot on the left showing a photon collider using the European XFEL $17.5$ GeV SC linac ($x=0.65$) and the plot on the right for an optimal optical laser set-up ($x=4.8$).
\begin{figure}[htbp]
    \centering
    \begin{subfigure}{0.5\linewidth}
        \centering
        \includegraphics[width=0.9\linewidth]{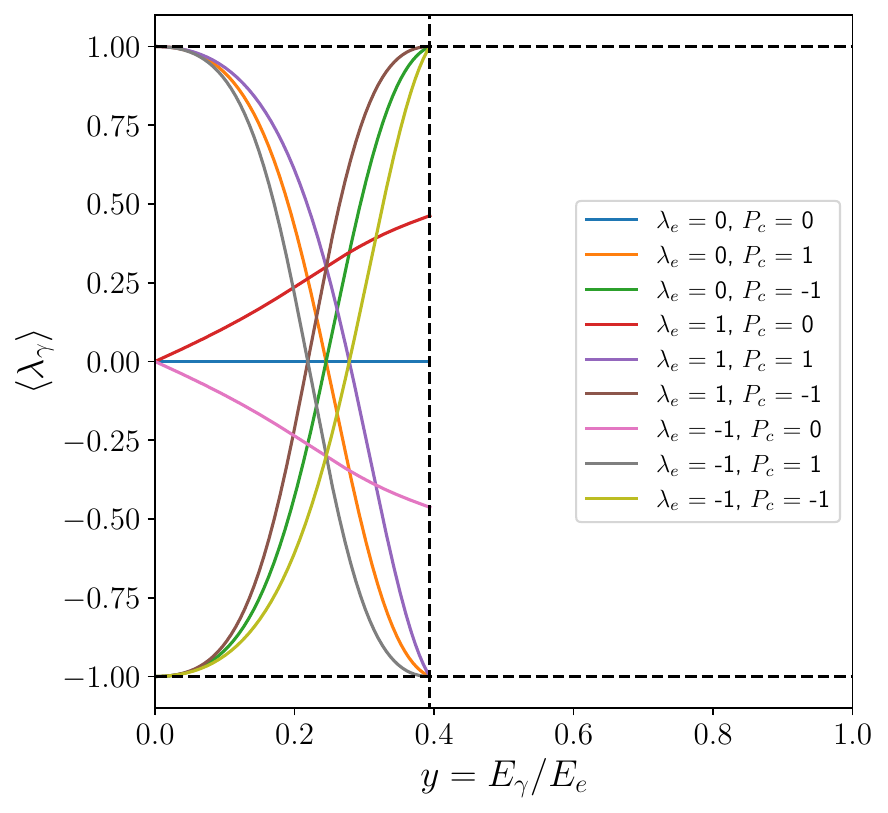}
    \end{subfigure}
    \begin{subfigure}{0.5\linewidth}
        \centering
        \includegraphics[width=0.9\linewidth]{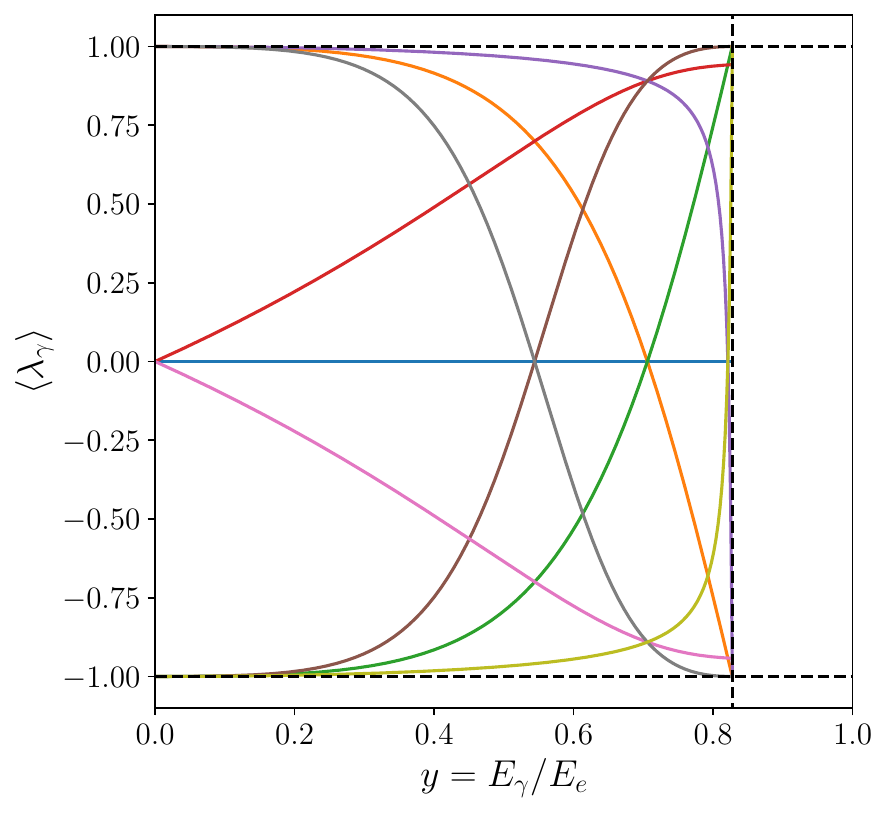}
    \end{subfigure}
    \caption{Mean helicity of Compton scattered photons with $x=0.65$ (left) and $x=4.8$ (right) for different initial laser photon polarisations $P_c$ and electron helicity $\lambda_e$.}
    \label{fig:helicitiesExample} 
\end{figure} 
We can hereby separate the resulting distributions into three groups: a) first for low $y:$ $\langle\lambda_\gamma\rangle=-1$ and for $y_{max}$ we get $\langle\lambda_\gamma\rangle=+1$ (brown, green and light-green curves); b) the opposite ones, starting for low $y$ at $\langle\lambda_\gamma\rangle=+1$ (grey, orange and purple curves) and c) three special ones (red, blue and pink curves) that start at $\langle\lambda_\gamma\rangle=0$ and either only get partially polarised or remain unpolarised when both electron and laser photon are unpolarised. Significant are the curves for which the electron beam is unpolarised but the photon beam is polarised (brown and pink) as these show that the backscattered photon is also polarised as well as that by flipping the laser photon polarisation from $P_c=-1$ to $P_c=+1$ or vice versa we can also completely flip the high energy photons polarisation distribution. When the electron beam is polarised too, in order to keep the overall spectrum and to only invert the polarisation behavior of the scattered photons, it is necessary to keep $\lambda_eP_c$ unchanged, therefore changing the polarisation of the laser as well as of the electron beam (orange and green curve).
Including the helicity of the final state photons separates the luminosity distribution into two states with total helicity of the colliding photons of $J_z=0\text{ and }2$, changing eq. (\ref{eq:lumiGamma}) into
\begin{equation}
    \begin{split}
        \frac{1}{k^2L_{ee}}\frac{\text{d}L_{\gamma\gamma}^{+\pm}}{\text{d}z} = &2z\int_{z^2/y_{max}}^{y_{max}}\frac{\text{d}y}{y} \left(\frac{1\pm\lambda_{\gamma,1}\lambda_{\gamma,2}}{2}\right) f(x,y)f\left(x,\frac{z^2}{y}\right) F(\rho),
    \end{split}
    \label{eq:lumiPOLARIZED}
\end{equation}
where $L^{+\pm}_{\gamma\gamma}$ represents $J_z=0$ for $++$ and $J_z=2$ for $+-$, $\lambda_{\gamma,i}$ are given by eq. (\ref{eq:helicityDistribution}) with the same values for $y$ as the $f(x,y)$ functions and unchanged $F(\rho)$ representing for simplicity the second and third line in Eq. (\ref{eq:lumiGamma}). The sum of both results in the total luminosity distribution as before $L_{\gamma\gamma}=\frac{1}{2}(L_{\gamma\gamma}^{++}+L_{\gamma\gamma}^{+-})$,, as shown in the right plot of Fig. \ref{fig:dLdz_polarized} where for different values of $\rho$ the total luminosity as well as the two  total helicity distributions are shown as well. For this choice of initial polarisations at high $z$ the spectrum is dominated by $J_z=0$ and with increasing $\rho$, the peak only decreases very slightly, while the tail nearly completely vanishes. The separation into these two distinct total helicity states is beneficial when searching for new particles or specific CP states \cite{Telnov_2020}.

\subsection{Nonlinear QED effects and multiple scattering}
\begin{figure}[htbp]
  \begin{subfigure}[b]{0.5\linewidth}
    \centering
    \includegraphics[width=0.85\linewidth]{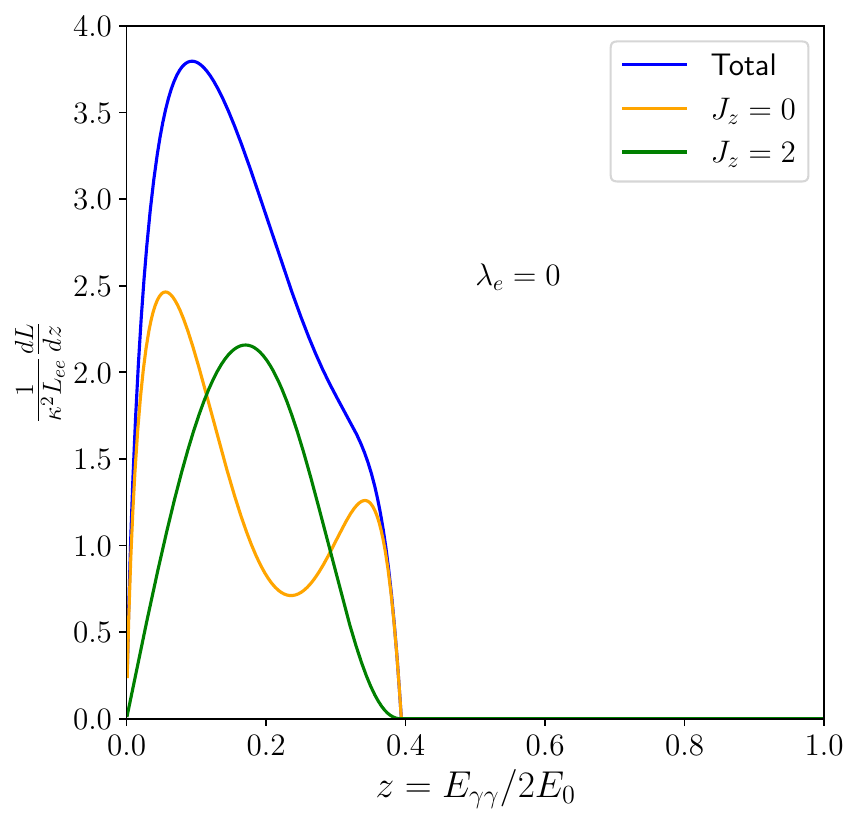} 
  \end{subfigure}
  \begin{subfigure}[b]{0.5\linewidth}
    \centering
    \includegraphics[width=0.85\linewidth]{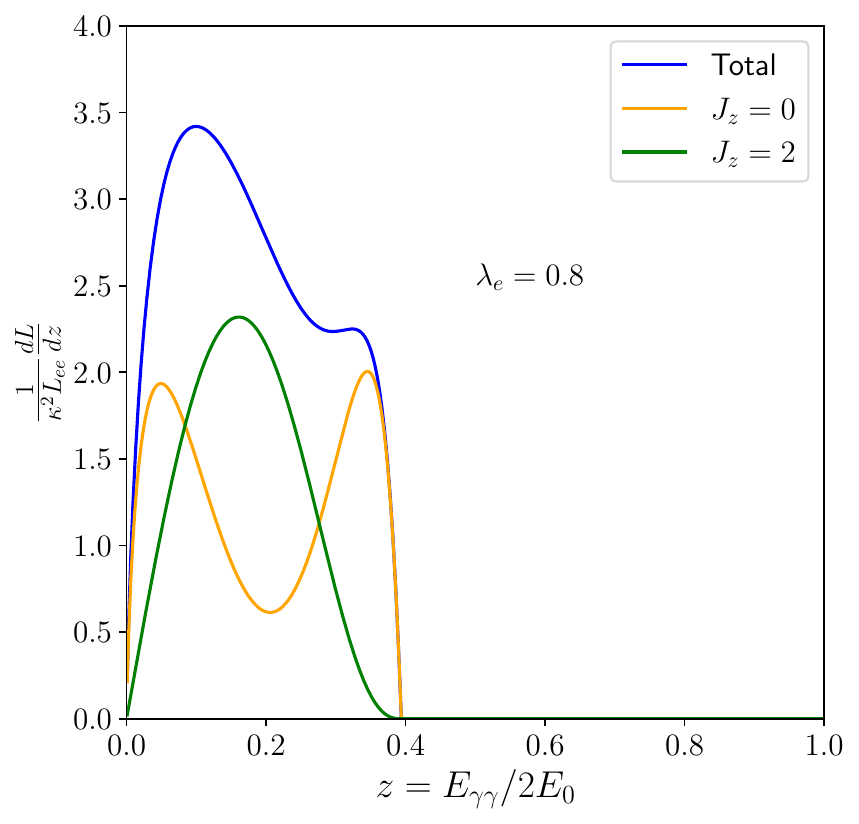}
  \end{subfigure} 
  \begin{subfigure}[b]{0.5\linewidth}
    \centering
    \includegraphics[width=0.85\linewidth]{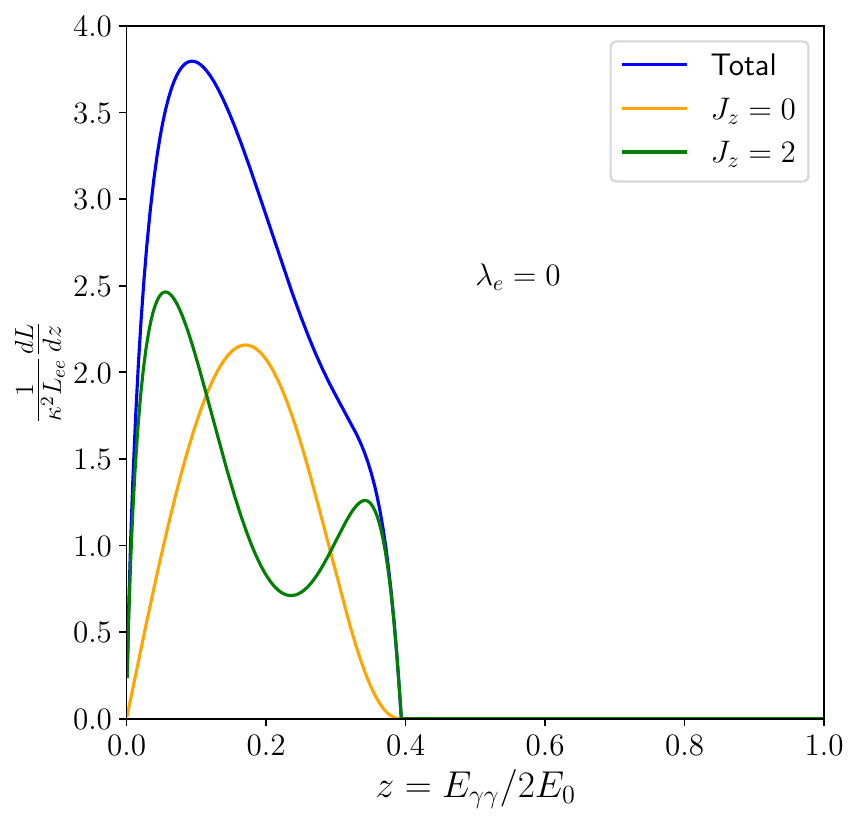}
  \end{subfigure}
  \begin{subfigure}[b]{0.5\linewidth}
    \centering
    \includegraphics[width=0.85\linewidth]{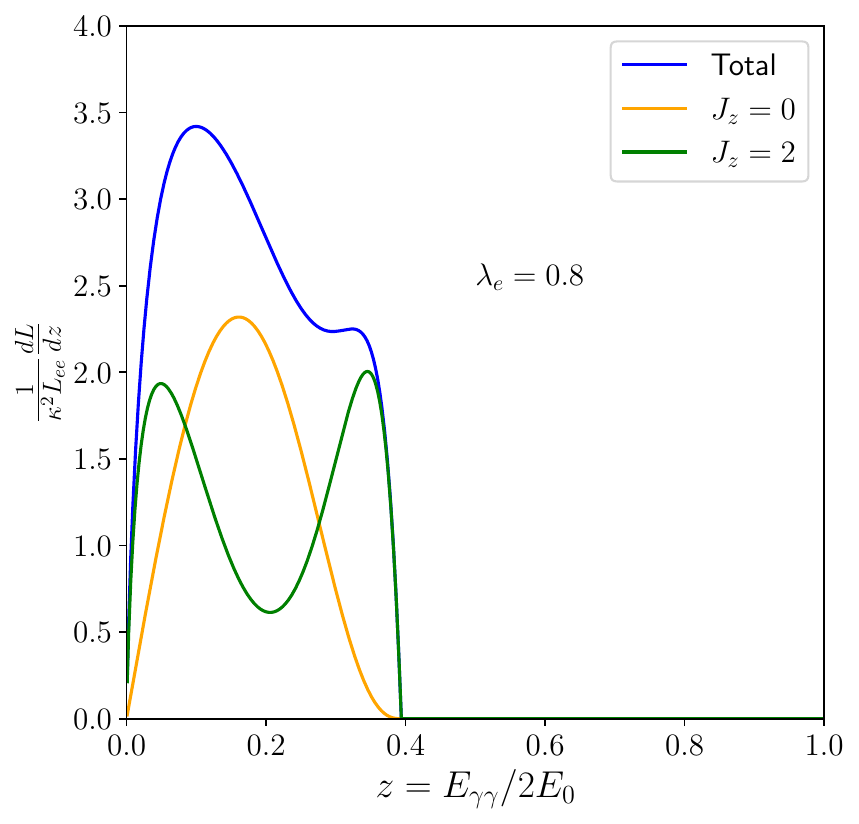}
  \end{subfigure} 
  \caption{Luminosity spectrum for a photon collider at the beam dump of the European-XFEL for unpolarised electrons (left panels) and longitudinal polarised electrons $\lambda_e=0.8$ ($80\%$) (right panels). The bottom two plots show the spectrum for the same setup and the same $\lambda_e P_c$, but on one beam side the electron polarisation and laser polarisation have been flipped.}
  \label{fig:final_analytical_spectra} 
\end{figure}

\begin{figure}[htbp]
    \centering
    \begin{subfigure}{0.5\linewidth}
        \centering
        \includegraphics[width=0.9\linewidth]{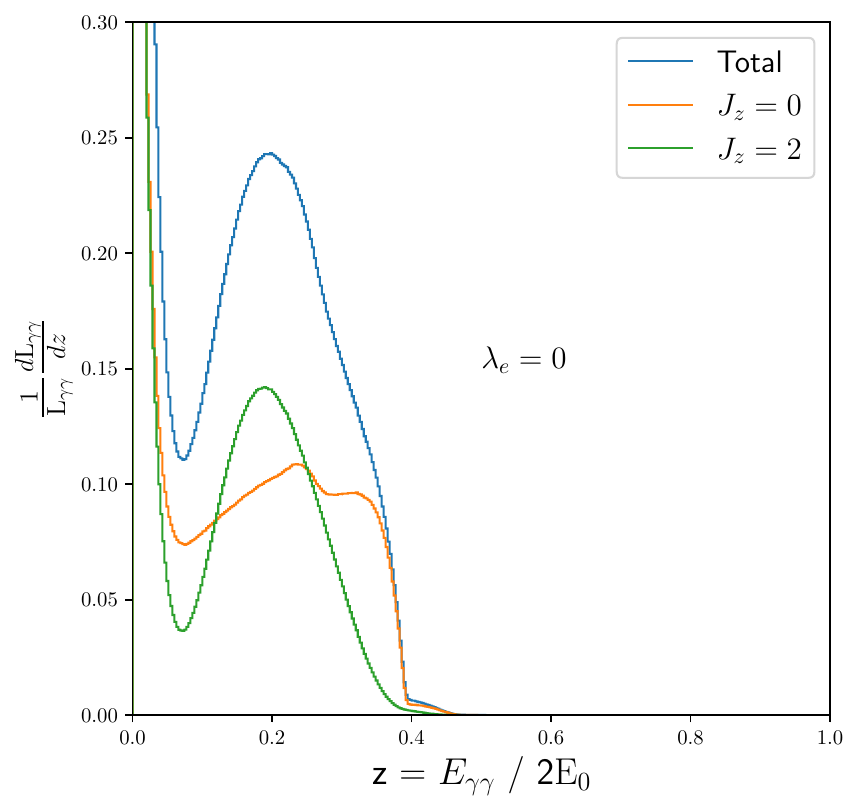}
    \end{subfigure}
    \begin{subfigure}{0.5\linewidth}
        \centering
        \includegraphics[width=0.9\linewidth]{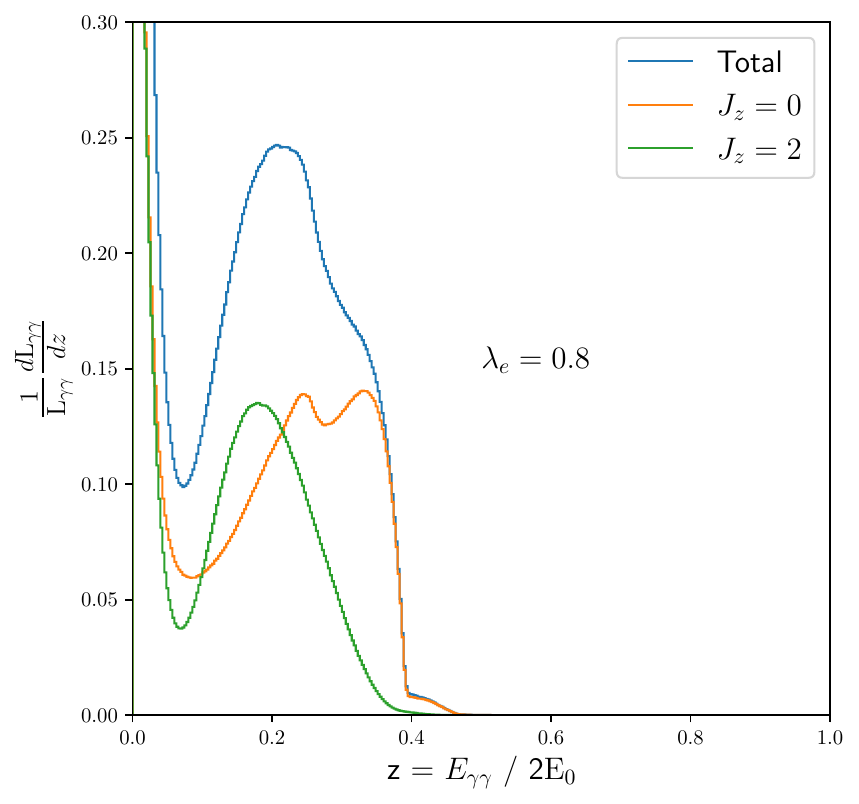}
    \end{subfigure}
    \caption{The realistic luminosity distribution for a photon collider for a) unpolarised electrons (left) and b) for an incoming electron beam polarisation with $\lambda_e=0.8$ (right). In both cases the laser photon polarisation is $P_c=-1$. The total spectrum is shown in blue and for the two polarisation states $J_z=0$ in red and $J_z=2$ in green.} 
    \label{fig:CAIN_spectrum}
\end{figure}

For a more realistic spectrum further factors have to be taken into consideration. Due to the very strong laser wave field in the conversion region, multiple laser photons can interact with a single electron simultaneously, known as nonlinear QED effects. These are expressed by the parameter \cite{TESLA_TDR}
\begin{equation}
    \xi^2 = \frac{e^2 \bar{F}^2 \hbar^2}{m_e^2 c^2 \omega_0^2} = \frac{2 n_\gamma r_e^2\lambda}{\alpha} = \left(\frac{\lambda}{2 \pi m_e}\right)^2 \mu_0 c P,
\end{equation}
where $\bar{F}$ is the r.m.s. strength of the electric field in the laser wave, $n_\gamma$ is the density of laser photons and $P$ is the laser power density. The nonlinear QED effects can be described as an effective increase of the electron mass $m_e^2\rightarrow m_e^2 (1+\xi^2)$, which then shifts the parameter $x$ of the photon collider $x\rightarrow x / (1 + \xi^2)$. Including this shift in eq. (\ref{eq:omega}) leads to a slight decrease in the maximum energy of the scattered photons,
\begin{equation}
    E_{\gamma,\text{max}}=\frac{x}{x+ 1+\xi^2} E_0.
\end{equation}
The energy spectrum of the Compton scattered photons and the luminosity distribution for different values of $\xi^2$ can be seen for example in Fig. 1.3.5 in \cite{TESLA_TDR}. The maximum of the luminosity spectrum (distribution) shifts to lower energies for higher values of $\xi^2$ and additional harmonics appear. 
Another important effect is multiple Compton scattering, where an electron scatters on further laser photons after the initial scattering. These electrons have lost energy, corresponding to the photon spectrum, and have an energy of as low as $E_1 \approx E_0-E_{\gamma,\text{max}}=E_0/(x+1)$ after one scattering process. Therefore a photon resulting from a secondary scattering process is softer, resulting in an increase in low energy photons with $y$ values close to 0.The analytical spectra of the 12 GeV photon collider not including $\xi^2$ and multi Compton scattering are shown in Fig. \ref{fig:final_analytical_spectra}, for unpolarised electrons (left) and $80\%$ longitudinal polarised electrons (right), as well as showing the effect that by keeping the product $\lambda_e P_c$ unchanged, while changing $\lambda_e$ and $P_c$ each allows one to invert the distribution of the polarised states $J_z=0$ and $J_z=2$. In the unpolarised electrons case this requires to only change one of the laser polarisation, while for the polarised case, both the electron polarisation and laser polarisation on one beam side have to be flipped.
The other effects can also be considered by further modifying the equations, but are better evaluated using Monte-Carlo simulation codes such as \texttt{CAIN} \cite{CHEN1995107}, which is able to generate realistic luminosity spectra for photon colliders.
Please note that in the case before we only considered the parameter $x$ and the electron beam energy $E_0$, but when using \texttt{CAIN}, one has to fully define the laser and electron beams, due to it using full beam simulations and including all multi particle effects.
Currently, \texttt{CAIN} supports spatial laser profiles of gaussian or donut form. In order to study a flat-top laser, we first create the laser profile following the mathematical description \cite{SLAC:1996mfa} and use that profile as our input for the photon collider.
The important parameters for the $e^-$- and laser-beams are given in Tab. \ref{tab:parameters}, taken from \cite{Telnov_2020} and references therein. The laser is not optimized for an electron collider at $32$ GeV (i.e. European-XFEL), but for a collider between $100$ and $1000$ GeV. Such a laser set-up is ideal to test the performance of and to understand all the technical challenges of  a photon collider. A laser with $x=4.8$ at $E_{ee}=35$ GeV would not necessarily offer a greater SM physics case and at the same time is very challenging if not currently impossible to achieve \cite{Telnov_2020}.\\
The resulting luminosity spectrum is shown in \cref{fig:CAIN_spectrum} for the case of an unpolarised electron beam (left) and with $80\%$ circular polarised electrons (right)\footnote{Plotting here in 100 MeV steps, compared to the 350 MeV steps in \cite{Telnov_2020}.}. It can be seen that this spectrum has some similarities to the spectra resulting from the analytical expressions, cf. \cref{fig:final_analytical_spectra}. The difference is mainly at lower energies, where \texttt{CAIN} also includes beamstrahlung and multiple scattering, therefore, has a far larger contribution for low $z$. At higher values of $z$ the shape follows the analytical patterns with a couple of extra bumps and slight changes in overall contribution. Still in both cases the $J_z=0$ state is dominant for large $z$ and depending on the need for the search can be optimised at what $z$ which of the two polarisation states is needed.
\begin{table}[htbp]
    \centering
    \begin{tabular}{|c|c|}
        \hline
        Electron beam energy $E_0$ [GeV] & 17.5 \\
        $\beta_x$/$\beta_y$ [$\mu$m] & 70/70 \\ 
        $\gamma\varepsilon_x$/$\gamma\varepsilon_y$ [mm$\cdot$ mrad] & 1.4/1.4 \\ 
        $\sigma_x$/$\sigma_y$ [nm] & 53/53 \\ 
        $\sigma_z$ [$\mu$m] & 70 \\
        Bunch charge [$10^{10}$ $e^-$] & 0.62 \\
        Collision rate [kHz] & 13.5 \\
        Crossing angle [mrad] & $\sim30$\\ 
        b ($C$-$IP$ distance) [mm] & 1.8 \\
        Laser photon $\lambda$ (energy) [$\mu$m (eV)] & 0.5 (2.33) \\
        Laser flash energy [J] & 3 \\
        Laser pulse duration [ps] & 2 \\
        f$\#\equiv F/D$ of laser system & 27 \\
        Parameter $x$ and $\xi^2$ & 0.65, 0.05 \\
        Power density laser [$10^{21}$W/m$^2$] & 1.4 \\
        $\alpha_{e\lambda}$ [mrad] & 47\\ 
        \hline
    \end{tabular}
    \caption{Parameters of the photon collider considered for $2 E_0=35$ GeV, based on the superconducting linac of the European-XFEL, taken from \cite{Telnov_2020} and references therein.}
    \label{tab:parameters}
\end{table}

\section{Light-by-Light scattering at the photon collider}
\label{sec:lbyl}
\begin{figure}[h]
    \centering
    \begin{subfigure}{0.32\textwidth}
        \centering
        \includegraphics[width=\linewidth]{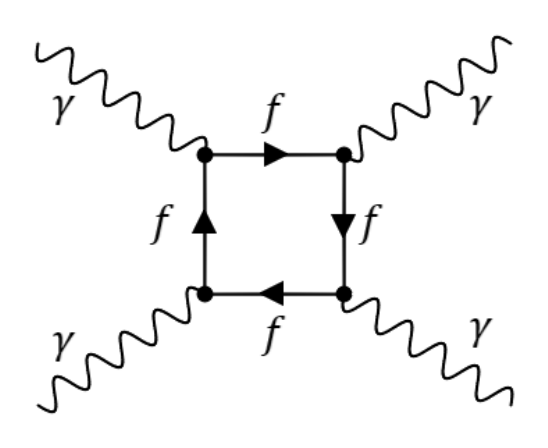}
    \end{subfigure}
    \begin{subfigure}{0.32\textwidth}
        \centering
        \includegraphics[width=\linewidth]{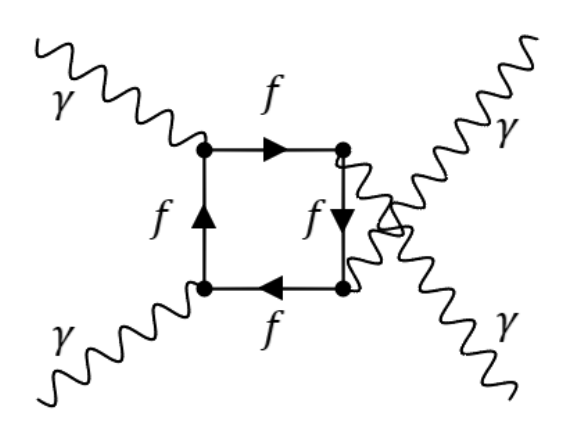}
    \end{subfigure}
    \begin{subfigure}{0.32\textwidth}
        \centering
        \includegraphics[width=\linewidth]{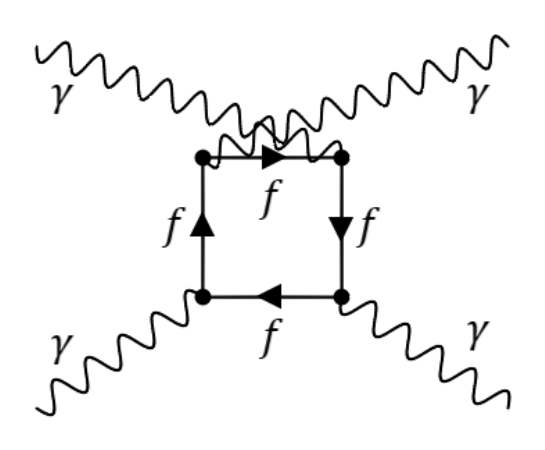}
    \end{subfigure}    
    \caption{The feynman diagrams of light-by-light scattering including fermions. Three further diagrams with opposite direction of momentum in the fermion loop are not shown.}
    \label{fig:FermionLbyL}
\end{figure}

\begin{figure}[htbp]
    \centering
    \includegraphics[width=0.7\textwidth]{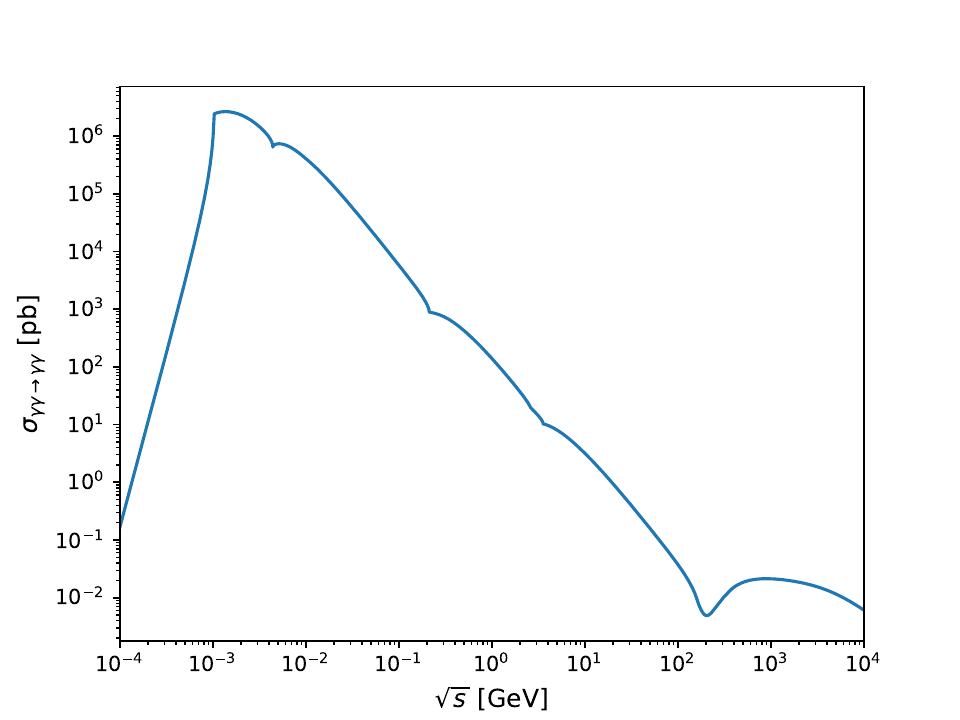}
    \caption{QED cross section of light-by-light scattering.}
    \label{fig:LbyL}
\end{figure}

\begin{figure}[htbp]
    \centering
    \includegraphics[width=0.7\textwidth]{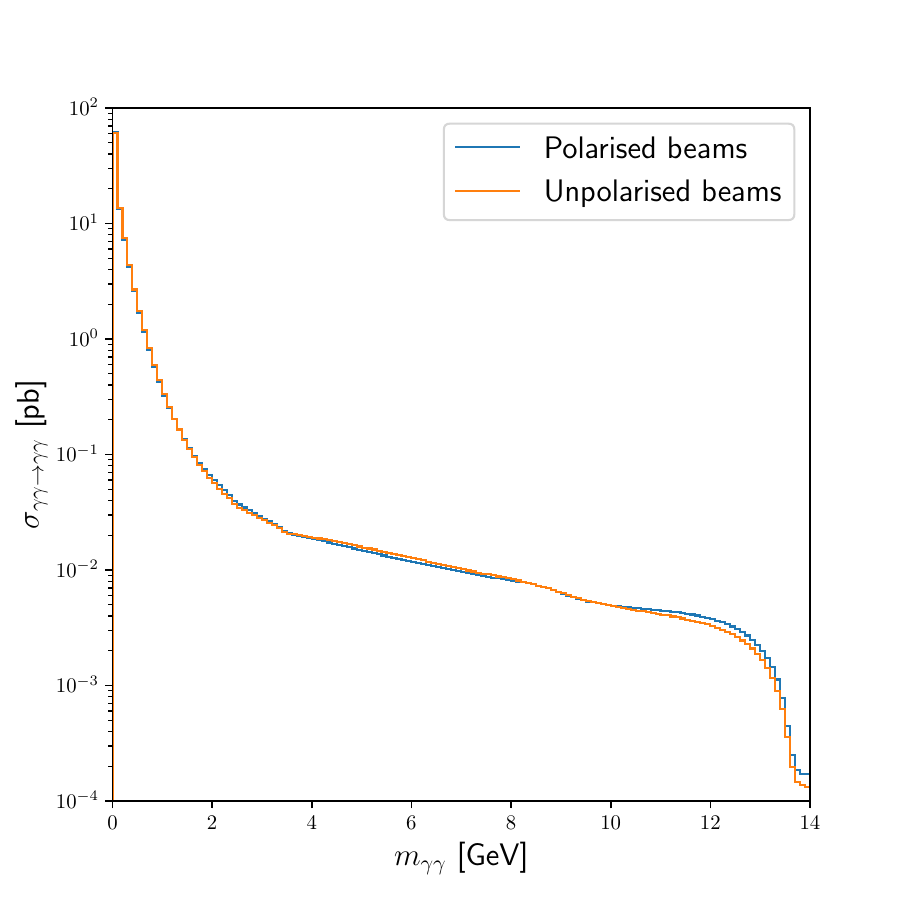}
    \caption{QED cross section of light-by-light scattering at a photon collider with energy $E_{\gamma\gamma}< 12$ GeV with unpolarised electron beams (orange curve) and $80\%$ circularly polarised electron beams. The spectrum simulated with \texttt{CAIN} with a bin size of 100 MeV.}
    \label{fig:Photon-collider-LbyL}
\end{figure}

A very important process in the SM that can be observed directly at a photon collider is the scattering of light-by-light $\gamma\gamma\rightarrow\gamma\gamma$, whose Feynman diagrams including fermions are shown in Fig. \ref{fig:FermionLbyL}. Here, only three of the total six non-identical fermionic diagrams are presented, the diagrams not shown only differ by the direction of the arrow in the loops.
Apart from fermionic contributions, the $W$-boson also contributes (the corresponding diagrams can be found in Figs. (2,3) in \cite{Boehm}). The latter ones are negligible at lower energies but are required to retain gauge invariance for completeness as well as for analyses at future high-energy colliders, these have been included in this study as well. 
In the center of mass frame the differential cross section for this $2\rightarrow 2$ process is given by
\begin{equation}
    \frac{\text{d}\sigma}{\text{d}\Omega} = \frac{1}{64\pi^2}\frac{1}{s}|M|^2.
    \label{eq:2to2CrossSection}
\end{equation}
It is useful to evaluate the squared amplitude of light-by-light scattering in terms of helicity amplitudes where each photon can take two independent helicity states $\lambda=\pm$, resulting in a total of $2^4=16$ possible invariant helicity amplitudes. These amplitudes are denoted as $M_{\lambda_1\lambda_2\lambda_3\lambda_4}$, where $\lambda_i$ represents the helicity state of the initial ($i=1,2$) and final ($i=3,4$) photon. Taking the symmetries of the process into account, the amount of independent amplitudes can be reduced. Due to $P$-invariance the initial number of amplitudes can be halved \cite{DeTollis:1964una,DeTollis:1965vna}
\begin{equation}
    \begin{split}
        M_{++++} = M_{----}, &\quad M_{--++} = M_{++--},\\
        M_{-+-+} = M_{+-+-}, &\quad M_{+--+} = M_{-++-},\\
        M_{---+} = M_{+++-}, &\quad M_{++-+} = M_{--+-},\\
        M_{+-++} = M_{-+--}, &\quad M_{-+++} = M_{+---}.
    \end{split}    
\end{equation}
Further making use of $T$-invariance they can be reduced to $5$ amplitudes \cite{DeTollis:1964una,DeTollis:1965vna}
\begin{equation}
    \begin{split}
        M_{+-++} = M_{+++-}, &\quad M_{++-+} = M_{-+++},\\
        M_{+---} = M_{--+-}, &
    \end{split}
\end{equation}
and lastly using crossing symmetry 
\begin{equation}
    \begin{split}
        M_{+-+-}(s,t,u) &= M_{++++}(u,t,s),\\
        M_{+--+}(s,t,u) &= M_{++++}(t,s,u),
        \label{eq:crossingsym}
    \end{split}
\end{equation}
where $s,t,u$ are the Mandelstam variables, $3$ independent helicity amplitudes to be evaluated are left
\begin{equation}
    M_{++++}(s,t,u), \quad M_{++--}(s,t,u), \quad M_{+++-}(s,t,u).
\end{equation}
In order to calculate the cross section for a specific initial and final photon polarisation, it is only necessary to replace $M$ with the respective amplitude, with regard to the symmetries above. When evaluating the unpolarised cross section one sums over the final photon polarisations and averages over the initial, replacing\footnote{Here note the extra factor $1/2$ compared to \cite{landau1982qed}.}
\begin{equation}
    |M|^2 \rightarrow \frac{1}{2}(|M_{++++}|^2+|M_{++--}|^2+|M_{+-+-}|^2+|M_{+--+}|^2+4|M_{+++-}|^2),
    \label{eq:SquaredAmplitudes}
\end{equation}
resulting in the unpolarised differential cross section\footnote{When inserting \cref{eq:SquaredAmplitudes} into \cref{eq:2to2CrossSection} another factor of $1/2$ needs to be included due to the identical particles in the final state. The final equation agrees with Eq. (5) \cite{Gounaris:1998qk}, but differs from Eq. (3) in \cite{Inan:2020aal} by a factor of 2.}
\begin{equation}
    \frac{d\sigma}{d\Omega} = \frac{1}{256\pi^2 s} (|M_{++++}|^2+|M_{+-+-}|^2+|M_{+--+}|^2+|M_{++--}|^2+4|M_{+++-}|^2).
    \label{eq:differentialHeli}
\end{equation}
The \textsc{Mathematica} package \texttt{FeynArts} \cite{FeynArts} was used to generate and evaluate these diagrams, further applying \texttt{FormCalc} \cite{FormCalc} allows us to choose the polarisations for the particles in the process, and to calculate the three independent helicity amplitudes.
The full SM amplitude can be split into a sum of bosonic and fermionic helicity amplitudes:
\begin{equation}
    M_{\lambda_1,\lambda_2,\lambda_3,\lambda_4}^{\text{SM}} = \sum_{fermion}M_{\lambda_1,\lambda_2,\lambda_3,\lambda_4}^{\text{fermion}} + M_{\lambda_1,\lambda_2,\lambda_3,\lambda_4}^{\text{boson}}.
\end{equation}
For the three amplitudes that contain fermions, we get:
\begin{align}
    M_{++++}^{\text{fermion}}(s, t, u) =& 8\alpha^2Q_f^4N_c \Bigg\{-1 + \frac{u-t}{s} \left[B_0(u,m_f)-B_0(t,m_f)\right] \nonumber \\
    &+ \left[ \frac{4m_f^2}{s}+2\left(\frac{tu}{s^2}-\frac{1}{2}\right) \right] \left[ uC_0(u,m_f)+tC_0(t,m_f) \right]  \label{eq:FERM++++}\\
    &-2m_f^2s\left( \frac{m_f^2}{s}-\frac{1}{2} \right)\Big[ D_0(s,t,m_f)+D_0(s,u,m_f)+D_0(t,u,m_f)\Big]   \nonumber \\
    &-tu\left( \frac{4M_f^2}{s}+\frac{tu}{s^2} -\frac{1}{2} \right) D_0(t,u,m_f)\Bigg\}, \nonumber
\end{align}
\begin{equation}
    M_{++--}^{\text{fermion}}(s, t, u) = 8\alpha^2Q_f^4N_c \Big\{ 1-2m_f^4\left[ D_0(s,t,m_f)+D_0(s,u,m_f)+D_0(t,u,m_f) \right]\Big\},
    \label{eq:FERM++--}
\end{equation}
\begin{align}
    M_{+++-}^{\text{fermion}}(s, t, u) &= 8\alpha^2Q_f^4N_c \Bigg\{ 1-m_f^2(s^2+t^2+u^2)\left[ \frac{1}{tu} C_0(s,m_f) + \frac{1}{su} C_0(t,m_f) + \frac{1}{st} C_0(u,m_f)\right] \nonumber \\
    &-m_f^2 \Bigg[ \left( 2m_f^2+\frac{st}{u} \right) D_0(s,t,m_f) + \left( 2m_f^2+\frac{su}{t} \right) D_0(s,u,m_f) \label{eq:FERM+++-} \\
    &+ \left( 2m_f^2+\frac{tu}{s} \right) D_0(t,u,m_f) \Bigg] \Bigg\}, \nonumber  
\end{align}
where $m_f$ is the mass of the fermion that has to be summed over at the end, $Q_f$ is the charge of the fermion, $N_c$ is the color factor (1 for leptons and 3 for quarks) and $B_0, C_0, D_0$ are the Passarino-Veltman scalar integrals \cite{Passarino:1978jh}. The bosonic contribution is
\begin{align}
    M_{++++}^{\text{boson}}(s, t, u) &= 12\alpha^2 \Bigg\{ 1- \frac{u-t}{s} \Big[ B_0(u,m_W)-B_0(t,m_W) \Big] \nonumber\\
    &-\left(\frac{4m_W^2}{s}+2\Big(\frac{tu}{s^2}-\frac{4}{3}\Big)\right) \Big[uC_0(u,m_W)+tC_0(t,m_W)\Big] \label{eq:W++++}\\
    &+\left(2m_W^2\Big(\frac{m_W^2}{s}-\frac{4}{3}\Big)\right) \Big[D_0(u,t,m_W)+D_0(s,u,m_W) \nonumber \\
    &+D_0(s,t,m_W)\Big]+ts\left(\frac{4m_W^2}{u}+\frac{ts}{u^2}-\frac{4}{3}\right) D_0(s,t,m_W)\Bigg\}, \nonumber
\end{align}
\begin{equation}
    M_{++--}^{\text{boson}}(s, t, u) = 12\alpha^2 \Big\{-1+2m_W^4 \Big[ D_0(u,t,m_W)+D_0(s,u,m_W)+D_0(s,t,m_W) \Big] \Big\},
    \label{eq:W++--}
\end{equation}
\begin{align}
    M_{+++-}^{\text{boson}}(s, t, u) &= 12\alpha^2 \Bigg\{ -1+m_W^2(s^2+t^2+u^2)\left[ \frac{1}{tu} C_0(s,m_W) + \frac{1}{su} C_0(t,m_W) + \frac{1}{st} C_0(u,m_W)\right] \nonumber \\
    &+m_W^2 \Bigg[ \left( 2m_W^2+\frac{st}{u} \right) D_0(s,t,m_W) + \left( 2m_W^2+\frac{su}{t} \right) D_0(s,u,m_W)  \label{eq:W+++-}\\
    &+ \left( 2m_W^2+\frac{tu}{s} \right) D_0(t,u,m_W) \Bigg] \Bigg\}. \nonumber
\end{align}
Inserting these coefficients in \cref{eq:FERM++++,eq:FERM++--,eq:FERM+++-,eq:W++++,eq:W++--,eq:W+++-} shows that two of the bosonic amplitudes are proportional to the fermionic amplitudes, when replacing the fermion mass by the $W$-boson mass,
\begin{equation}
    M_{++--}^\text{boson}(s,t,u)=-\frac{C_b}{C_f}\left\{ M_{++--}^\text{fermion}(s,t,u); m_f\rightarrow m_W \right\},
\end{equation}
\begin{equation}
    M_{+++-}^\text{boson}(s,t,u)=-\frac{C_b}{C_f}\left\{ M_{+++-}^\text{fermion}(s,t,u); m_f\rightarrow m_W \right\}.
\end{equation}
Including all these contributions, the full QED light-by-light scattering cross section $\sigma_{\gamma\gamma\rightarrow\gamma\gamma}$ can be seen in Fig. \ref{fig:LbyL}. Each bump corresponds to a new fermion becoming on-shell, and at lower energies $\sqrt{s} < 2 m_W$ the overall shape is dominated by the electron contribution. The other fermions cause only a small shift. For energies $\sqrt{s} > 2m_W$, the cross section is mainly given by the $W$-boson contributions and the fermionic contributions are negligible.
The highest value for the cross section is achieved just when the electrons become on-shell, while at higher energies the cross section decreases, but is around $10$ pb at about $\sqrt{s}=100$ GeV, when the $W$-boson contribution dominates. At the proposed gamma gamma collider with $E_{\gamma\gamma} < 12$ GeV, the cross section is still of order $\mathcal{O}$(pb), promising great prospects for their detection and detailed analyses. 
As shown in the previous chapter, the luminosity distribution for a photon collider is not a delta peak but depends strongly on the parameters of the electron beam and the laser. 
In order to get the full cross section for the linear collider system, one has to weight the differential cross section in the photon system d$\sigma_{\gamma\gamma}$ with the normalized luminosity spectrum,
\begin{equation}
    \text{d}\sigma = \int_{0}^{z_{max}}\text{d}z \int_{z^2/y_{max}}^{y_{max}}\text{d}y\frac{1}{L_{\gamma\gamma}}\frac{\text{d}L_{\gamma\gamma}}{\text{d}z}\text{d}\sigma_{\gamma\gamma\rightarrow\gamma\gamma}.
    \label{eq:GammaGamma}
\end{equation}
For simplicity, a luminosity spectrum with $\rho=0$, cf. Eq. \ref{eq:rho}, is commonly used. When also considering the scattered photon polarisation, see \cref{eq:lumiPOLARIZED}, the unpolarised cross section can be calculated by $\sigma=\frac{1}{2}(\sigma^{+-}+\sigma^{++})$.
When including the polarisation of the scattered photons the cross section is given by
\begin{equation}
    \text{d}\sigma^{+\pm} = \int_{0}^{z_{max}}\text{d}z \int_{z^2/y_{max}}^{y_{max}}\text{d}y\frac{1}{L_{\gamma\gamma}^{+\pm}}\frac{\text{d}L_{\gamma\gamma}^{+\pm}}{\text{d}z}\text{d}\sigma_{\gamma\gamma\rightarrow\gamma\gamma}^{+\pm},
    \label{eq:GammaGamma_polarised}
\end{equation}

where the polarised photon-photon cross section only includs the 8 amplitudes with the incoming polarisation

\begin{align}
    |M^{++}|^2 &= \frac{1}{2}(|M_{++++}|^2+|M_{++--}|^2+2|M_{+++-}|^2),\\
    |M^{+-}|^2 &= \frac{1}{2}(|M_{+-+-}|^2+|M_{+--+}|^2+2|M_{+++-}|^2).
    \label{eq:SquaredAmplitudes_polarised}
\end{align}
Using \cref{eq:GammaGamma_polarised} with the luminosity spectra simulated with \texttt{CAIN}, the resulting SM light-by-light scattering cross section is shown in \cref{fig:Photon-collider-LbyL}. The spectrum for polarised beams has a slightly higher contribution at higher energies. This can also be seen in the total cross section at increased energies $E_{\gamma\gamma}>10$ GeV. In the range of $E_{\gamma\gamma}<10$ GeV, there is no noticeable difference between the cross section for polarised and unpolarised electron beams, making this process accessible at the proposed collider without additional electron polarisation. For the photon collider with a luminosity spectrum of unpolarised electron beams, the total light-by-light scattering cross section is $97.04$ pb. We chose an energy cut of 100 MeV excluding soft photons and chose $|\theta|\geq 5^{\circ}$ for this study.  With the same setup but electron beams that are $80\%$ circularly polarised the total cross section is $96.20$ pb, due to the fact that the spectrum is shifted to higher energies.

\section{ALPs production at a photon collider with Energy $E_{\gamma\gamma}< 12$ GeV}
\label{sec:alp}
\begin{figure}[h]
    \centering
    \begin{subfigure}{0.32\textwidth}
        \centering
        \includegraphics[width=\linewidth]{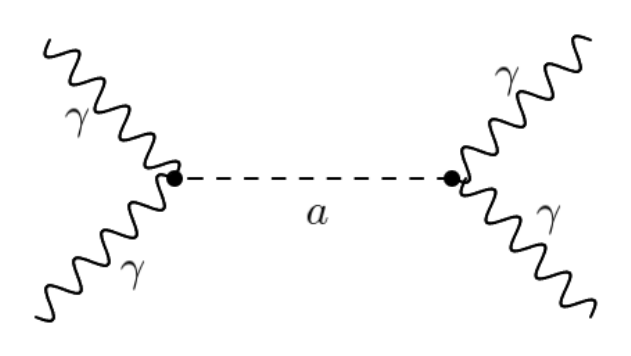}
    \end{subfigure}
    \begin{subfigure}{0.32\textwidth}
        \centering
        \includegraphics[width=\linewidth]{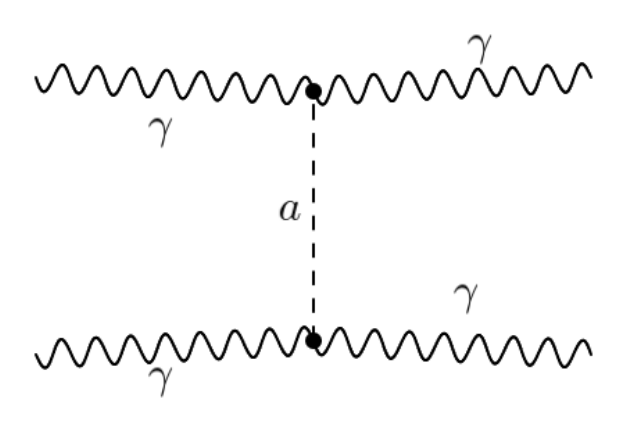}
    \end{subfigure}
    \begin{subfigure}{0.32\textwidth}
        \centering
        \includegraphics[width=\linewidth]{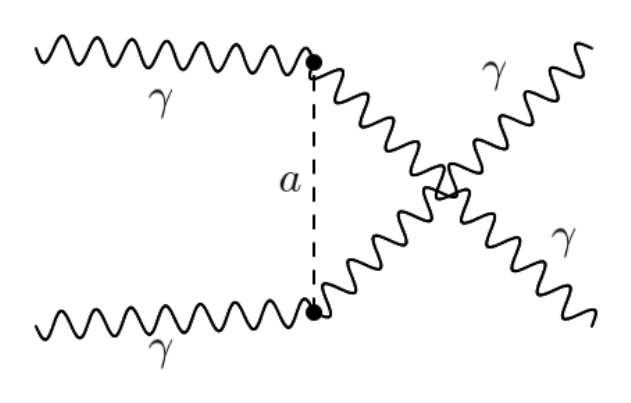}
    \end{subfigure}    
    \caption{Feynman diagrams for ALP-production in the $\gamma\gamma\rightarrow\gamma\gamma$ channel.}
    \label{fig:ALPproduction}
\end{figure}
\begin{figure}[htbp]
    \centering
    \begin{subfigure}{0.5\linewidth}
        \centering
        \includegraphics[width=0.9\linewidth]{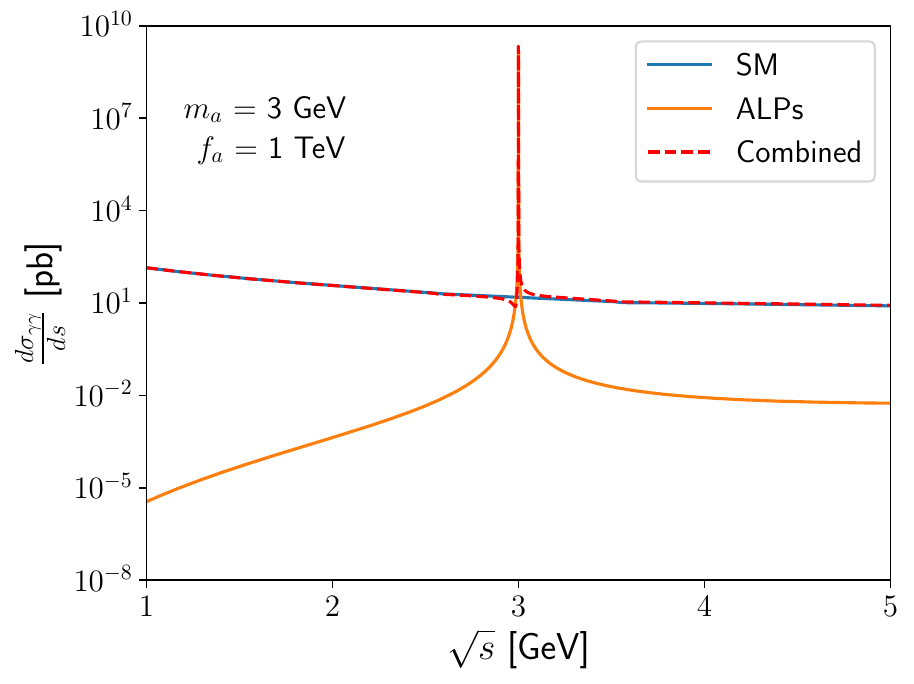}
    \end{subfigure}
    \begin{subfigure}{0.5\linewidth}
        \centering
        \includegraphics[width=0.9\linewidth]{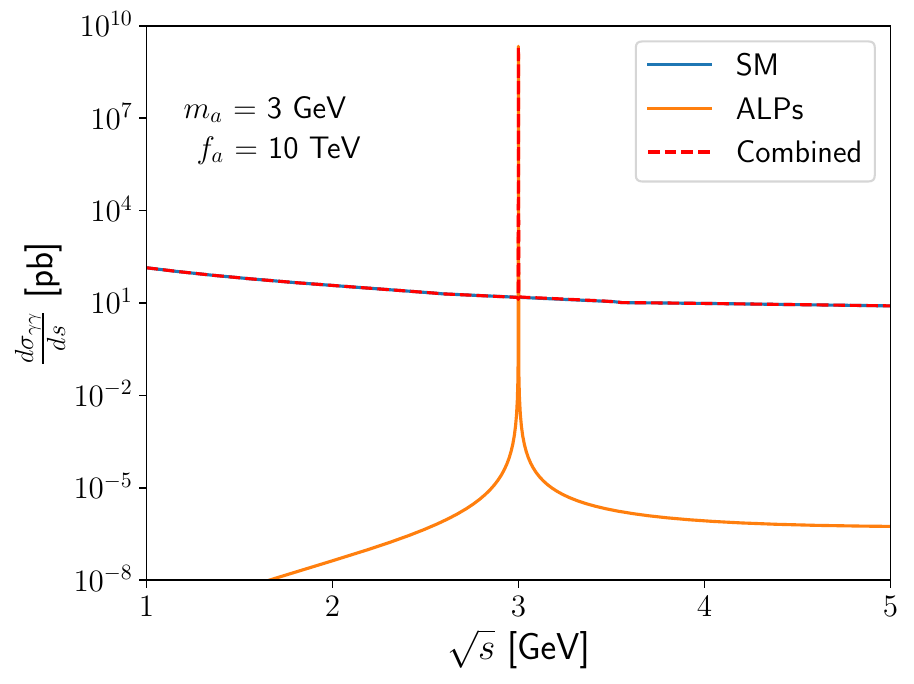}
    \end{subfigure}
    \begin{subfigure}{0.5\linewidth}
        \centering
        \includegraphics[width=0.9\linewidth]{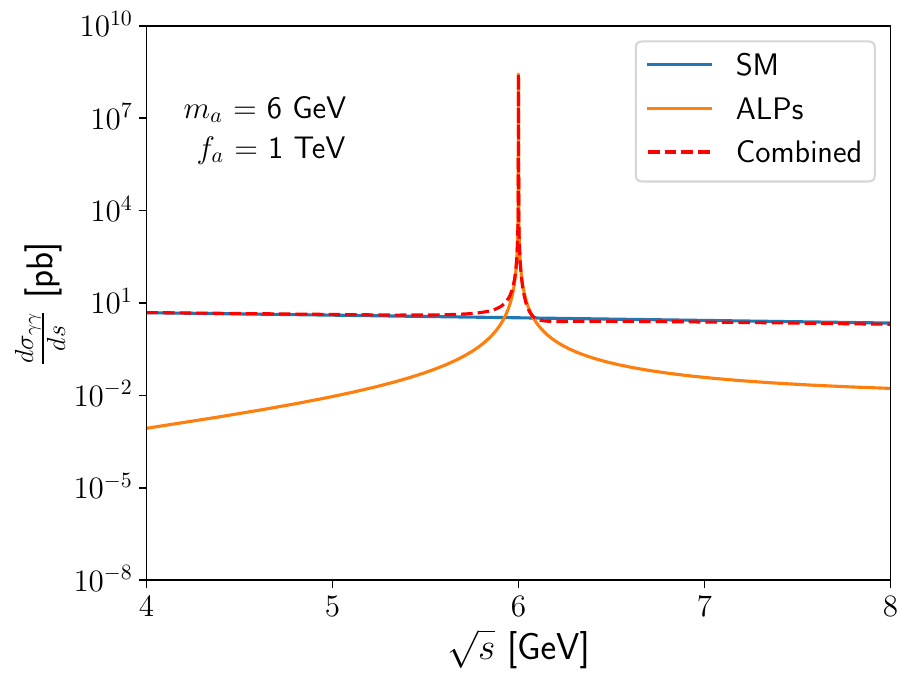}
    \end{subfigure}
    \begin{subfigure}{0.5\linewidth}
        \centering
        \includegraphics[width=0.9\linewidth]{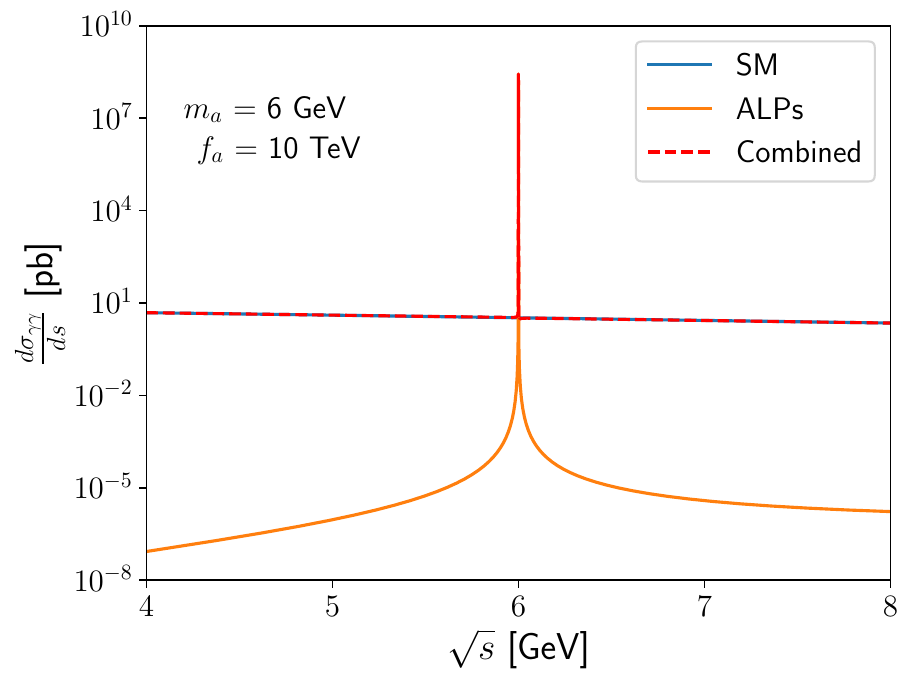}
    \end{subfigure}
    \caption{Differential cross section for light-by-light scattering, for the cases of only SM contribution, only ALPs contribution and the combined cross section, for $m_a=3$ GeV (upper plots) and $m_a=6$ GeV (lower plots), $f_a=1$ TeV (left plot), $f_a=10$ TeV (right plot) and BR $= 1$.} 
    \label{fig:ALPs_inGammaGamma}
\end{figure}
\begin{figure}[htbp]
  \begin{subfigure}[b]{0.5\linewidth}
    \centering
    \includegraphics[width=0.85\linewidth]{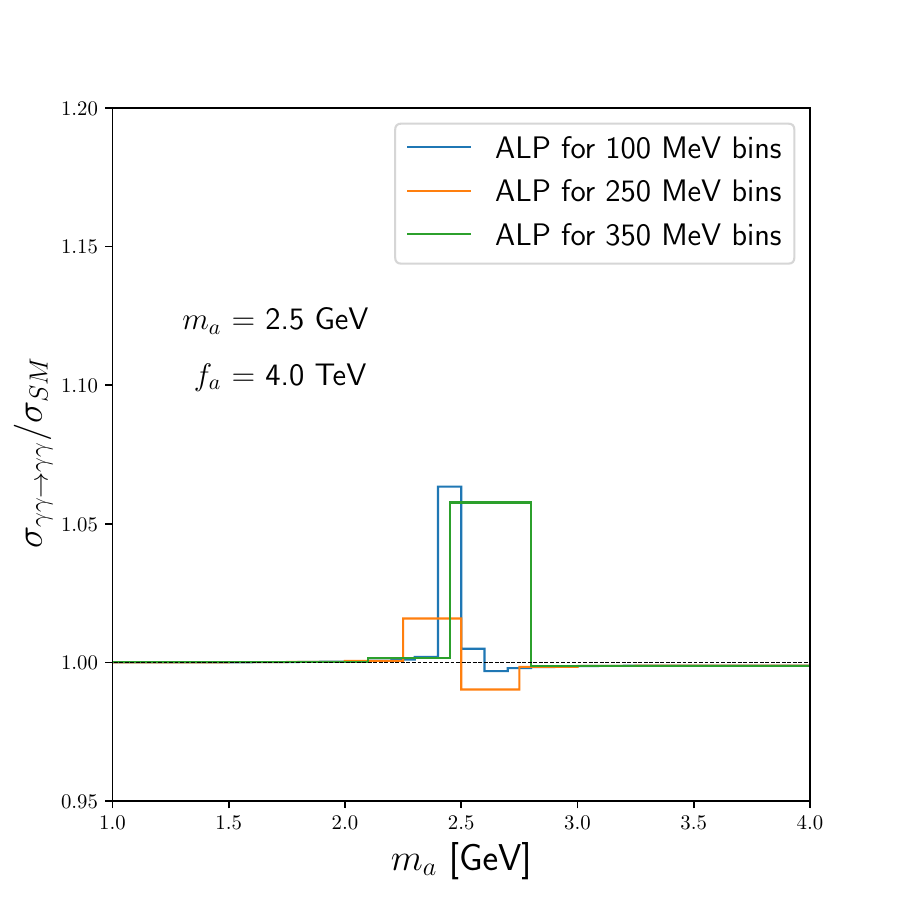} 
  \end{subfigure}
  \begin{subfigure}[b]{0.5\linewidth}
    \centering
    \includegraphics[width=0.85\linewidth]{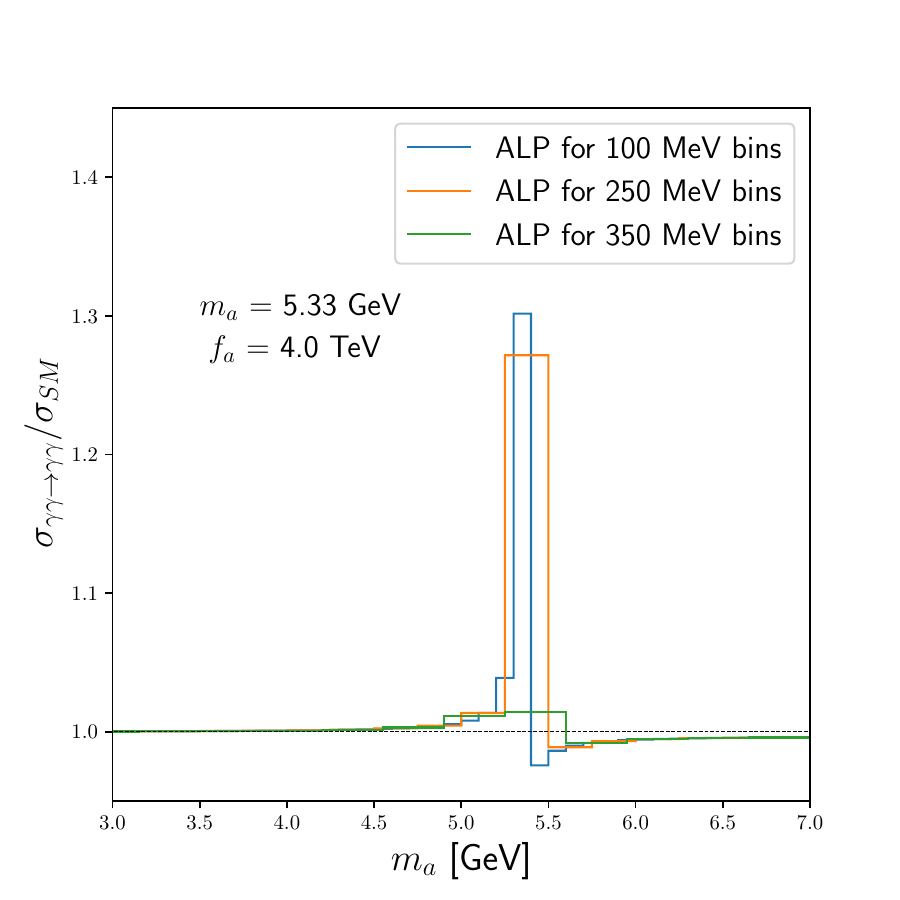}
  \end{subfigure} 
  \begin{subfigure}[b]{0.5\linewidth}
    \centering
    \includegraphics[width=0.85\linewidth]{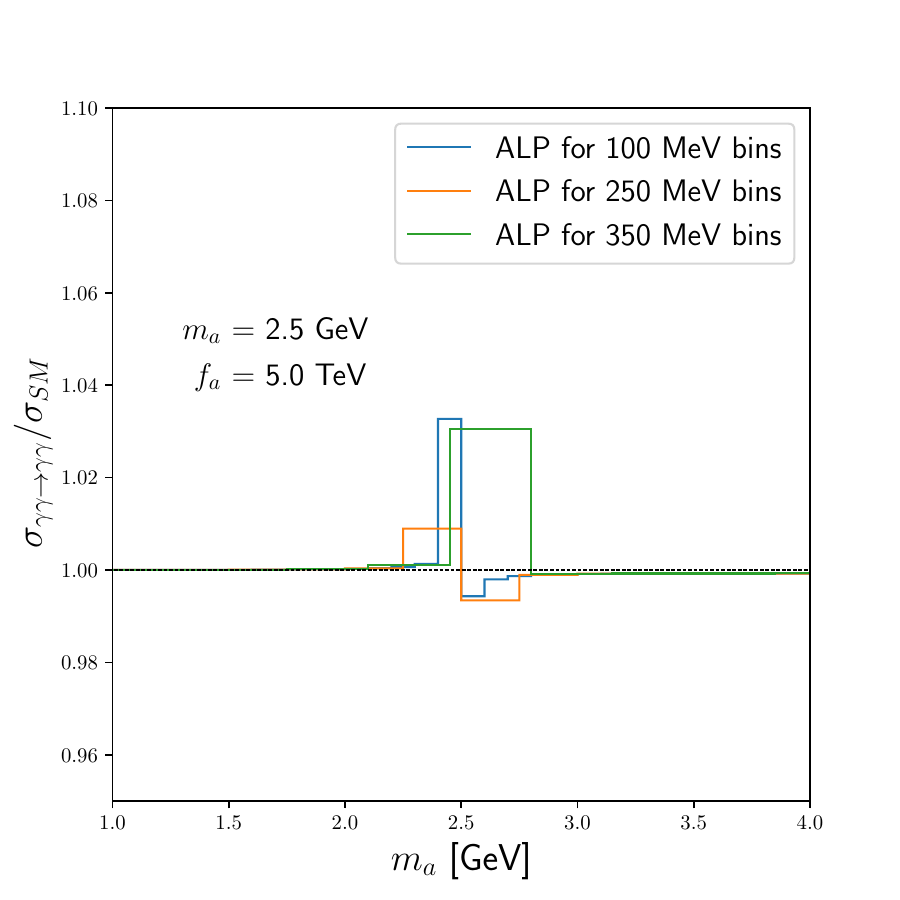}
  \end{subfigure}
  \begin{subfigure}[b]{0.5\linewidth}
    \centering
    \includegraphics[width=0.85\linewidth]{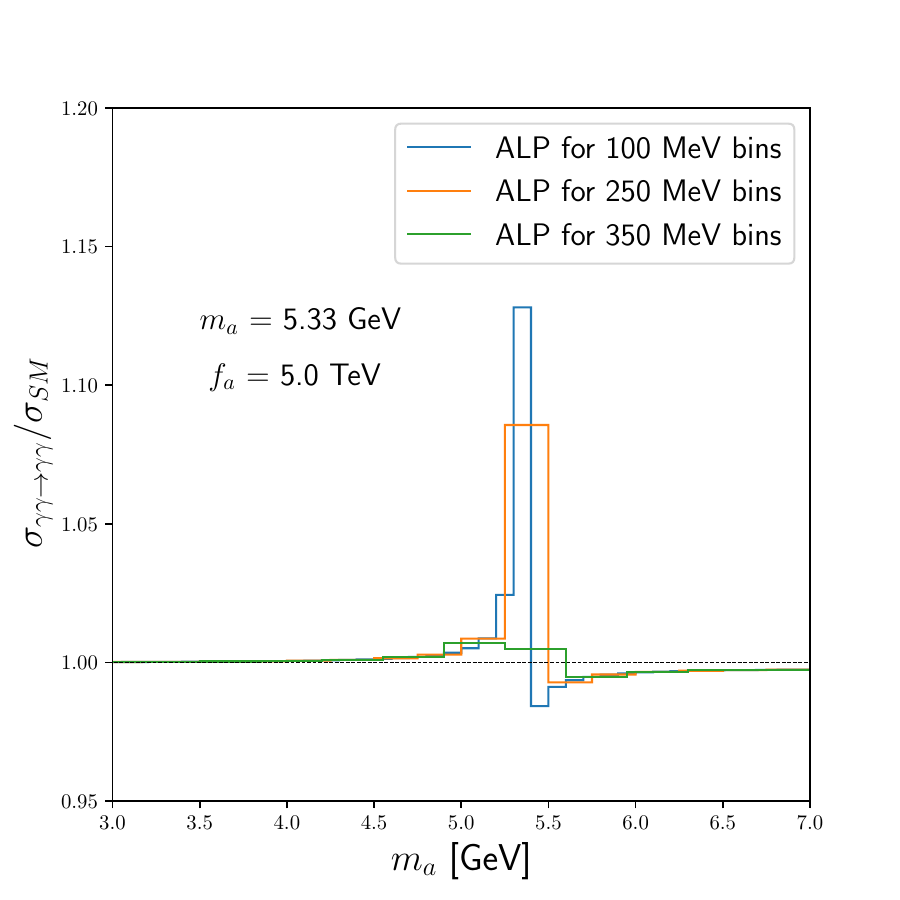}
  \end{subfigure} 
  \caption{Photon collider cross section for the process $\gamma\gamma\rightarrow a\rightarrow\gamma\gamma$ normalised over the SM light-by-light scattering cross section for different choices of bin sizes. The upper plots are for values $f_a=4$ TeV and the lower once for $f_a=5$ TeV. The left plots show for an ALPs mass $m_a=2.5$ GeV and the right plots for an ALPs mass $m_a=5.33$ GeV.}
  \label{fig:CAIN_ALP_bin_analysis} 
\end{figure}
\begin{figure}[htbp]
    \centering
    \begin{subfigure}{0.5\linewidth}
        \centering
        \includegraphics[width=0.9\linewidth]{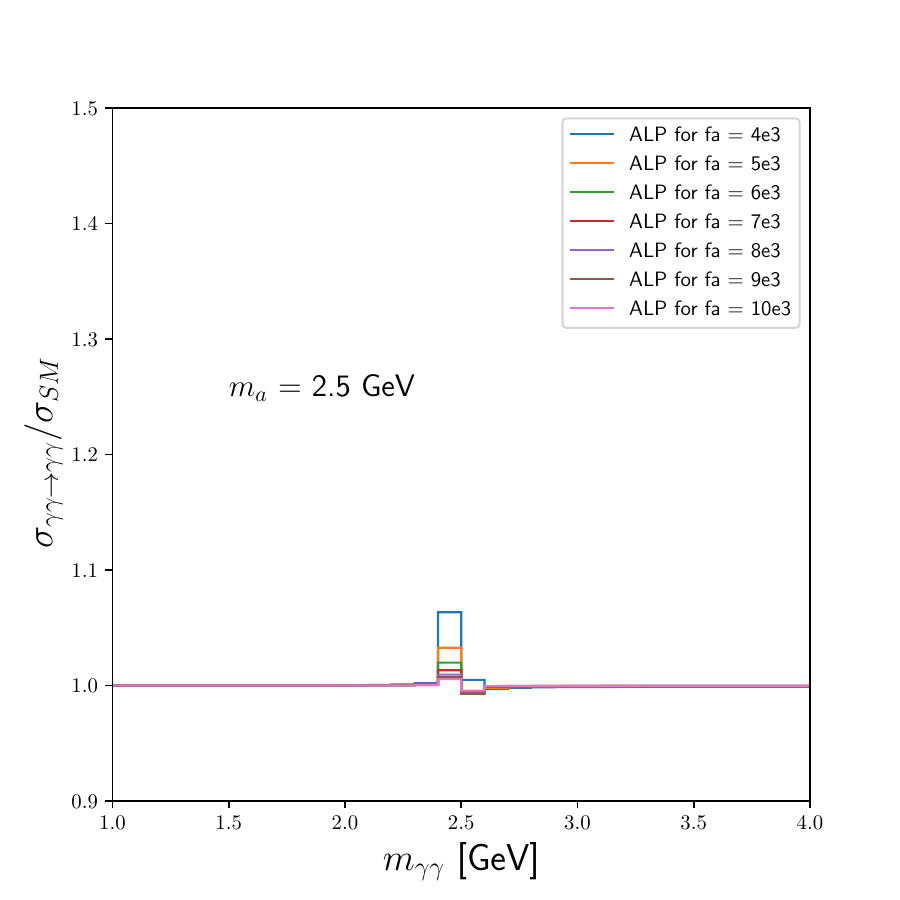}
    \end{subfigure}
    \begin{subfigure}{0.5\linewidth}
        \centering
        \includegraphics[width=0.9\linewidth]{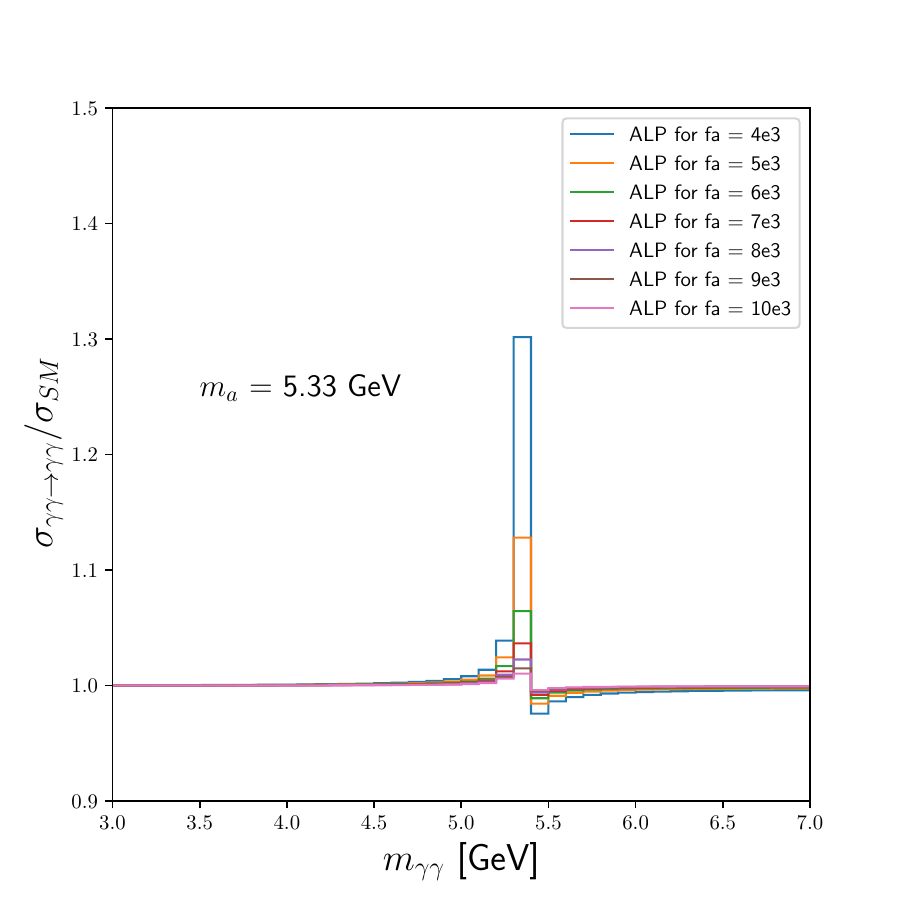}
    \end{subfigure}
    \caption{Photon collider cross section for the process $\gamma\gamma\rightarrow a\rightarrow\gamma\gamma$ normalised over the SM light-by-light scattering cross section for different values of $f_a$. Left plot shows an ALP mass of $m_a=2.5$ GeV and the right plot shows an ALP mass $m_a=5.33$ GeV.}
    \label{fig:ALPs_inGammaGamma_different_couplings}
\end{figure}

The Lagrangian describing the coupling of ALPs to photons at energies below the electroweak scale can be written as
\begin{equation}
    \mathcal{L}_{ALPs}=\frac{1}{2}\partial^\mu a \partial_\mu a-\frac{1}{2}m_a^2 a^2-\frac{a}{f_a}F_{\mu\nu}\Tilde{F}^{\mu\nu},
\end{equation}
where $a$ is the axion field, $m_a$ is the mass of the axion, $F_{\mu\nu}$ is the electromagnetic field strength tensor (with $\Tilde{F}^{\mu\nu}=\frac{1}{2}\epsilon^{\mu\nu\rho\sigma}F_{\rho\sigma}$) and $(f_{a})^{-1}$ the ALP-photon coupling constant\footnote{We note here the choice of the coupling constant $1/f_a$, following \cite{Inan:2020aal}. Another common choice is $\mathcal{L}_{ALPs}\subset -\frac{g_a}{4}aF_{\mu\nu}\Tilde{F}^{\mu\nu}$, with the relation $g_a=\frac{4}{f_a}$.}. 
Similar to the SM light-by-light scattering we can write the helicity amplitudes for the process $\gamma\gamma\rightarrow a\rightarrow\gamma\gamma$, cf. the Feynman diagrams in Fig. \ref{fig:ALPproduction}:
\begin{align}
    M_{++++}^{ALP}(s,t,u) &= \frac{4}{f_a^2} \frac{s^2}{s-m_a^2+im_a \Gamma_a},\label{eq:ALP++++}\\
    M_{++--}^{ALP}(s,t,u) &= \frac{4}{f_a^2} \left(\frac{s^2}{s-m_a^2+i m_a \Gamma_a} + \frac{t^2}{t-m_a^2} + \frac{u}{u-m_a^2} \right),\label{eq:ALP++--}\\
    M_{+-+-}^{ALP}(s,t,u) &= M_{++++}^{ALP}(u,t,s)= \frac{4}{f_a^2} \frac{u^2}{u-m_a^2},\label{eq:ALP+-+-}\\
    M_{+--+}^{ALP}(s,t,u) &= M_{++++}^{ALP} (t,s,u) = \frac{4}{f_a^2} \frac{t^2}{t-m_a^2},\label{eq:ALP+--+}\\
    M_{+++-}^{ALP}(s,t,u) &=0,\label{eq:ALP+++-}             
\end{align}
with the ALPs width
\begin{equation}
    \Gamma_a=\frac{1}{\text{Br}(a\rightarrow\gamma\gamma)}\frac{m_a^3}{4\pi f_a^2}.
    \label{eq:width}
\end{equation}
The ALP width $\Gamma_a$ has only been included in eq. (\ref{eq:ALP++++}) and in the first term of eq. (\ref{eq:ALP++--}) due to its negligible impact for energies away from $s\sim m_a^2$.
The amplitudes can then be inserted in eq. (\ref{eq:differentialHeli}), correspondingly, to obtain the complete $\gamma\gamma\rightarrow\gamma\gamma$ scattering amplitude
\begin{equation}
    M_{\lambda_1\lambda_2\lambda_3\lambda_4}=M^{ALP}_{\lambda_1\lambda_2\lambda_3\lambda_4}+M^{SM}_{\lambda_1\lambda_2\lambda_3\lambda_4}.
\end{equation}
In Fig. \ref{fig:ALPs_inGammaGamma} two example scenario for the masses $m_a=3$ GeV and $m_a=6$ GeV at two values of the coupling coefficient $f_a=1,10 \text{ TeV}$ and $BR=1$ are shown, comparing the SM light-by-light scattering cross section with the contribution from ALPs and with the total combined cross section of $\text{SM} + \text{ALP}$. The SM cross section is in the lenergy range ($E_{\gamma\gamma} \leq 5$ GeV) still of $\mathcal{O}(10)$ pb, while slightly below $10 \text{ pb}$ for ($E_{\gamma\gamma} > 5 \text{ GeV}$), including the very narrowly peaked ALP contribution around its resonance. The slight decrease in the SM background is clearly visible when looking at the ALP contribution and how much the peak (orange curve) sits above the SM cross section. In the case of the $m_a=3 \text{ GeV}$ ALP only a small delta peak reaches above the SM cross section and the combined shows a slight change from the SM curve, but for the $m_a=6 \text{ GeV}$ ALP the orange curve clearly goes beyond the SM cross section and also the combined cross section mimics the full form of the peak.
For small values of $f_a$ the dip-peak structure, for $m_a=3 \text{ GeV}$,  of the resonance is still visible, but at $f_a=10$ TeV, it is already barely visible. Furthermore, the actual peak increasingly takes the form of a delta peak for both masses. A narrow resonance would be completely lost when integrating over a large range of $\sqrt{s}$, even if the ALP increased the cross section by several orders of magnitude on resonance. 
The SM contributions to light-by-light scattering are $\propto 1/\sqrt{s}$ and as shown in the lower plots of Fig. \ref{fig:ALPs_inGammaGamma} at higher $m_a$, the ALP contribution increases compared to the SM contribution, giving a wider and more dominant peak for the same coupling value $f_a$.  
Due to the narrow peak the total cross section would be very close to the SM-like results. Therefore, we are studying the structure around the peak for the collider analysis. \\
The form of the peak depends on the size of the bins for the luminosity spectrum and are shown in \cref{fig:CAIN_ALP_bin_analysis} for two different ALP masses $m_a=2.5 \text{ and } 5.33$ GeV  and two values for the coupling $f_a=4, 5$ TeV, normalised to the SM light-by-light scattering cross section. The choice of the ALPs masses is due to different bin size choices where $m_a$ could be on the boundary between two bins. These two choices provides the possibilities just to analyse suitable scenarios for both cases. For a bin size of $350$ GeV only one bin differs from the SM cross section, for an ALP with mass $m_a=5.33$ GeV, however, more bins differ, but only at and below the $\%$ level. With a bin size of $250$ GeV at least two bins differ and a peak dip structure can be seen, one that is even clearer for a bin size of $100$ MeV: at the peak a change of $30\%$ ($13\%$) can be seen, for a coupling of $4$ TeV ($5$ TeV). Even the $250$ MeV bin size still gives an increase of close to $30\%$ for $m_a=5.33$ GeV and $f_a=4$ TeV (upper right plot). 
Taking the bin size of $100$ MeV and comparing different values for the coupling $f_a$, for the higher ALP mass, $m_a=5.33$ GeV a small deviation from the SM can be seen, at high value $f_a=10$ TeV (see the right plot Fig. \ref{fig:ALPs_inGammaGamma_different_couplings}) and for the lower mass $m_a=2.5$ GeV the deviation is below $10\%$ for any value of $f_a$. The difference between the masses at the same coupling comes from the fact that the SM background is larger at smaller energies $m_{\gamma\gamma}$ overtaking the shape of the ALPs peak.
The current bounds on the photon-ALPs coupling $(f_a)^{-1}$ in the range of $100\text{ MeV }< m_a < 5\text{ GeV}$ are at $f_a >4$ TeV cf. Fig. 9 in \cite{CMS:2024bnt}
as well as expected results from data being currently taken show $f_a >40$ TeV limits. Such a photon collider would offer an alternative check of the mass and ALP-photon-coupling, independent of other ALP couplings to for example electrons. There are several ways to optimism the analysis even further as, for example, taking the angle distribution of the SM background and the ALP signal events, but the principle is the same.

\section{Conclusion}
\label{sec:conclusion}
We have studied the unique advantages of extending future $e^+e^-$ linear colliders with a photon collider via Compton backscattered photons, giving access to $\gamma\gamma$ and $\gamma e$ interactions. The concept has been discussed for many years. Using an already existing electron beam like at the European XFEL would allow us to realise such a collider in a shorter time frame and with a cheap investment, developing the expertise and the knowledge as well as a prototype of the concept for any future collider facilities. We discussed different aspects that are of importance when considering the photon collider and the importance of using Monte-Carlo codes like \texttt{CAIN} to get the most realistic spectrum for any analysis. \\
This possible set-up has been used to calculate the light-by-light scattering cross section in the SM. We show example scenarios for light-by-light scattering with an additional ALP in the process. Using polarised electron beams only gives a slight advantage for processes at energies $E_{\gamma\gamma}>10$ GeV in this case, while below that energy there is no noticeable difference. The photon collider with energies $E_{\gamma\gamma}<12$ GeV is able to differentiate between pure SM and additional ALP in the mass range $1\text{ to }6$ GeV beyond current exclusion limits and independent of other ALP-particle couplings. \\
This photon collider would not just be a great opportunity to test the photon collider principles but offers already a rich SM and BSM phenomenology.\\

\noindent\textit{\textbf{Acknowledgements:}}
We would like to thank V. I. Telnov and K. Yokoya for many interesting discussions and
providing input for the usage of CAIN.
This study was supported by the Deutsche
Forschungsgemeinschaft (DFG, German Research Foundation) under MO-2197/2-1.
We also acknowledge support 
by the Deutsche Forschungsgemeinschaft (DFG, German Research Foundation) under Germany's Excellence Strategy --- EXC 2121 ``Quantum Universe'' --- 390833306.

\bibliographystyle{JHEP}
\bibliography{bibliography}

\end{document}